\documentclass[aps,amsfonts,amsmath,prd,preprint,nofootinbib]{revtex4}
\newcommand{\beq}{\begin{equation}}
\newcommand{\eeq}{\end{equation}}

\usepackage{csquotes}
\numberwithin{equation}{section}

\usepackage[title]{appendix}
\usepackage{verbatim}
\usepackage{subcaption}

\def\beq{\begin{equation}} 
\def\eeq{\end{equation}}
\def\beqa{\begin{eqnarray}}
\def\eeqa{\end{eqnarray}}
\def\p{\partial}

\usepackage{hyperref} 
\hypersetup{ 
	colorlinks=true, 
	linkcolor=black, 
	filecolor=black, 
	citecolor = blue,       
	urlcolor=blue, 
} 
\usepackage{graphicx}

\begin{document}

\title{Global String Instantons}
	
	%\author{Jose J. Blanco-Pillado, Georgios Fanaras and Alexander Vilenkin}
	
\author{Jose J. Blanco-Pillado,}
\affiliation{Department of Physics, University of Basque Country, UPV/EHU, 48080, Bilbao, Spain}
\affiliation{EHU Quantum Center, University of Basque Country, UPV/EHU, 48080, Bilbao, Spain}
\affiliation{IKERBASQUE, Basque Foundation for Science, 48011, Bilbao, Spain}

\author{Georgios Fanaras,}
\author{Alexander Vilenkin,}
\affiliation{Institute of Cosmology, Department of Physics and Astronomy, Tufts University, Medford, MA 02155, USA}
	%\address{ Institute of Cosmology, Department of Physics and Astronomy,\\ 
	%	Tufts University, Medford, MA 02155, USA}

	\begin{abstract}
		
	%We study the formation of global strings during inflation through quantum mechanical tunneling. The instanton that describes the nucleation process is a two-parameter function of the string core thickness and its gravitational backreaction on the the background deSitter spacetime. We manage to obtain the instanton solutions for the complete range of these parameters by carrying out a numerical integration via shooting methods. The computational techniques are non-trivial in the sense that the integration boundaries exhibit numerical singularities and as such we have to resort to multiple shooting techniques. Our results are in-line with previous findings on the nucleation of topological defects, that is that after a certain threshold of the string core thickness and/or its gravitational backreaction the configuration becomes homogeneous \`a la Hawking-Moss. We note that the two instantons are a natural continuation of eachother and once the inhomogeneous glboal string channel appears it will dominate the nucleation process regardless of the relative magnitude of its Euclidean action to its homogeneous counterpart.  

We study the formation of gravitating global strings through quantum mechanical tunneling. The instantons that describe the nucleation process are characterized by two parameters: the string core thickness and its gravitational backreaction controlled by the string core energy density. We obtain solutions across a wide range of these parameters by carrying out numerical integration via multiple shooting methods. Our results are in agreement with previous findings on the nucleation of other topological defects; specifically, after reaching a certain threshold for the string core thickness or its gravitational backreaction, the configuration becomes homogeneous in a manner akin to Hawking-Moss solutions. Additionally, we analyze the global structure of the analytical continuation of the solutions to Lorentzian signature, revealing the emergence of a region of spacetime that describes an anisotropic universe. Finally, we also discuss the relevance of these instantons in the context of quantum cosmology.

	\end{abstract}

	\maketitle

\tableofcontents

%\section{Comments}

%\textbf{I think the skeleton of the draft is ok. The phrasing might not be the best so maybe you can help with that. We need to add more references (e.g. R. Gregory and your relevant work). We should comment in the discussion about the AdS case for future work. I would argue that the main thing that has to be taken care of is the language. }

\section{Introduction}

Inflation is widely regarded as the leading paradigm in early universe cosmology \cite{Guth:1980zm, Linde:1981mu}. According to this picture the universe underwent a period of exponential expansion driven by a scalar field, the inflaton. This mechanism is responsible for producing on large scales a nearly homogeneous, isotropic and flat universe and generating the seeds for structure formation\footnote{For a review of the predictions of inflation see, for example, \cite{Tegmark:2004qd}.}. One of the problems that the inflationary scenario also addresses is the absence of magnetic monopoles in our local universe. According to Grand Unified Theories (GUTs) \cite{Georgi:1974sy, Georgi:1974yf}, magnetic monopoles should exist \cite{tHooft:1974kcl,Polyakov:1974ek} and be in abundance, arising as a result of phase transitions in the early universe \cite{, Zeldovich:1978wj,Preskill:1979zi}. However, this conclusion would be dramatically altered if a period of inflation occurred after the phase transition that produced the monopoles. The exponential expansion during this inflationary phase would dilute the monopole density to negligible levels making them effectively irrelevant for cosmology.

Monopoles belong to a wide class of objects called topological defects like domain walls and cosmic strings \cite{Kibble:1976sj,Vilenkin:1984ib}. Similarly to the case of monopoles, many scenarios involving domain walls are stringently constrained by cosmological observations \cite{Zeldovich:1974uw}. In contrast, cosmic string networks produced during phase transitions do not face such severe issues and have been extensively studied over the years in connection to their cosmological observables \cite{Vilenkin:2000jqa}. Most models explored in the literature consider the formation of these cosmic strings during a post-inflationary epoch, when the temperature of the universe is of the order of the energy scale characteristic of the objects being formed. This leads to the formation of a network
of defects that subsequently evolves in an expanding universe.
Owing to their topological stability, some of these objects could potentially survive until the present day. Even in cases where they decay, they may leave detectable imprints on certain cosmological observables, providing indirect evidence of their existence.

However, the formation of topological defects in a cosmological setting is not limited to post-inflationary mechanisms. It has been shown that monopole-antimonopole pairs, circular string loops, and spherical domain walls can also be produced during the inflationary stage through quantum mechanical tunneling \cite{Basu:1991ig, Basu:1992ue, Basu:1993rf}. The continuous production of such defects during inflation could lead to appreciable number densities in the present universe, provided their energy scale is not much higher than that of the expansion rate during inflation. The exponential expansion of the universe during inflation would stretch these defects to sizes that could extend well beyond the current horizon, depending on the time of their formation. In general, such processes result in a distribution of objects with varying sizes, which, as in previous scenarios, could also produce observable effects.

One of the goals of this paper is to study the instanton solutions that describe the nucleation of global strings during inflation. The core ideas underlying the process of string nucleation have been already presented in the literature. In particular, here we will follow the analysis of thick topological defect instantons described in \cite{Basu:1992ue}. The case of domain walls is analyzed there numerically both in the case of a fixed spacetime background and also taking into account gravitational back-reaction. In the case of strings, the authors only studied the analytic case of a fixed spacetime background when the defect solution approaches a homogeneous configuration i.e. its core thickness approaches the size of the Hubble horizon. In this work, we aim to extend the previous analysis by conducting a comprehensive numerical investigation of nucleating global strings with different core sizes, incorporating the effects of gravitational backreaction on the geometry of spacetime.

Global strings have been studied in the context of post-inflationary formation scenarios for several decades now \cite{global-strings}. Their relevance has significantly increased in recent years, as they provide a mechanism for generating a cosmological density of axion-like particles, which are potential candidates for dark matter. In these models, a network of global (axionic) strings forms during the early universe following a symmetry-breaking phase transition the so-called Peccei-Quinn transition. One key distinction between global strings and their local counterparts is their coupling to the Goldstone mode, which results in significant radiation of those massless particles (the would be axions) as the strings undergo relativistic motion.

Axionic string networks have two significant implications: first, they create a background of axion particles, and second, they also lead to the production of a gravitational wave background. While certain details of these scenarios remain under debate, the hope of these investigations is to be able to put robust constraints on the scale of the new physics that leads to the formation of strings. Moreover, some models predict the formation of primordial black holes via the collapse of the string network, triggered by a subsequent phase transition that creates walls bounded by the strings, a scenario that is also under current investigation \cite{Ferrer:2018uiu,Dunsky:2024zdo}.

In contrast, much less attention has been given to scenarios where topological defects are produced via quantum nucleation. Notable work has been done in this area for domain walls, particularly concerning the formation of primordial black holes and wormholes \cite{Garriga:1992nm,Garriga:2015fdk,Deng:2016vzb}. Our study marks a preliminary step in order to explore similar phenomena in the context of global strings. As we will discuss further, the spacetime structure induced by global string loops exhibits similarities to certain wormhole solutions found in previous studies of cosmological domain walls.

Another interesting aspect of our solutions is their interpretation as instantons within the framework of 
quantum cosmology. The results presented in this paper suggest that these instantons can be used to understand the nucleation of compact universes with a global string embedded in their geometry. The presence of the global string alters the symmetry of the solution compared to the standard symmetric de Sitter instanton. By analytically continuing these instantons to a Lorentzian signature, they can be identified as a natural way to describe the creation of an anisotropic universe from nothing.

The rest of the paper is organized as follows. In Sec 2 we illustrate how cosmic strings can be nucleated in de Sitter space through quantum mechanical tunneling in the thin wall approximation. We proceed in Sec 3 by identifying a possible field theory description of these type of instantons in the context of global strings. In Sec 4 we obtain instanton solutions for a global string in a background spacetime with positive cosmological constant $\Lambda=H^{2}>0$, while also incorporating the gravitational deformations to the geometry.  We numerically solve the instanton system of equations for different values of the string core thickness and the gravitational backreaction. In Sec 5 we describe the global structure of the
solutions as well as their interpretation. In Sec 6. we evaluate the string instanton action and compare it with the corresponding homogeneous configuration. We finally discuss our results in Sec 7.

\section{Quantum Nucleation of Thin Cosmic Strings}

%\subsection{Nambu-Goto approach}

In this section, we will lay out a heuristic picture for the nucleation of strings in a fixed background spacetime with positive vacuum energy. Our main goal is to motivate the rigorous and complete analysis that will follow in the subsequent sections \footnote{Here we follow closely the analysis presented in \cite{Basu:1991ig}.}. Therefore in order to simplify the problem we will restrict ourselves to the thin wall description of the string. Furthermore, we will also disregard for the time being any gravitational backreaction of any object present in our geometry and consider a fixed background given by the static representation of the de Sitter space metric, namely,
\beq
ds^{2}=-f(r)dt^{2}+\frac{dr^{2}}{f(r)}+r^{2}d\Omega^{2}~,
\label{dsmetric}
\eeq
where $f(r)=1-H^{2}r^{2}$, $d\Omega^{2}$ is the metric on a unit 2-sphere 
%with polar and azimuthal angles $\theta$, $\phi$ 
and $\Lambda=H^{2}$ the positive cosmological constant.

As we mentioned earlier, we begin by modeling the string using the Nambu-Goto action, which assumes the string can be effectively approximated by an infinitely thin relativistic object characterized solely by its tension, or energy per unit length. Namely we will consider the action,
\beq
S_{NG}=-\mu\int d^{2}\zeta\sqrt{-\gamma}~,
\eeq
where $\mu$ denotes the string tension and $\gamma$ is the determinant of the two dimensional worldsheet metric parametrized by the coordinates $\zeta$.

Taking these elements into consideration, we can now examine a circular loop
of string of radius $r=R(t)$ in the de Sitter coordinate system presented in Eq. (\ref{dsmetric}).
%This loop can be described by a worldsheet parametrized by the 
%coordinates $\zeta^a =(t, \theta)$, with the induced metric expressed as follows,
%\beq
%{ds_{2}}^{2}=\left(-f(R)+\frac{\dot{R}^{2}}{f(R)}\right)dt^{2}+R^{2}d\theta^{2}
%\label{2metric}~.
%\eeq
Assuming this type of loop, we can see that the Nambu-Goto action for this configuration becomes,
\beq
S_{NG}=-2\pi\mu\int dtR\sqrt{f(R)-\frac{\dot{R}^{2}}{f(R)}}~.
\eeq

This reduced action defines a Lagrangian and a conjugate momentum $p_{R}$ corresponding to the single degree of freedom in this simplified model, the radius $R(t)$. These quantities are defined as follows:
\beq
L=-2\pi\mu R\sqrt{f-\frac{\dot{R}^{2}}{f}}~~~~,~~~~p_{R}=\frac{2\pi\mu\dot{R}R}{f\sqrt{f-\frac{\dot{R}^{2}}{f}}}~.
\label{mom}
\eeq
In order to study the loop dynamics it will be  convenient to use the Hamiltonian constraint of the system:
\beq
E=p\dot{R}-L= \text{const.}
\label{Ham}
\eeq
Inserting the expression for the momentum, the above conservation law recasts to:
\beq
\dot{R}^{2}+V(R)=0~,
\label{rdotv}
\eeq
where the effective potential is given by, $V(R)=-f(R)^{2}+R^{2}f(R)^{3}\epsilon^{-2}$ and $\epsilon=E/(2\pi\mu)$ is the ratio of the string energy to its tension. This can be viewed as the equation describing the classical motion of a particle of zero energy moving in the potential $V$.

It can be shown that for $2H\epsilon<1$, $V(R)$ forms a potential barrier and we can have bounded classical trajectories. The turning points of the potential can be readily found as the solutions to $V(R)=0$:
\beq
R_{1,2}=\sqrt{\frac{1}{2H^{2}}\left(1\pm\sqrt{1-4\epsilon^{2}H^{2}}\right)}~.
\eeq
The trajectories of (\ref{rdotv}) can be classified as follows. The string loop can start from zero size at $R=0$ expand to $R=R_{1}$, bounce and recollapse. An alternative solution is a contracting loop that bounces at $R=R_{2}$ and then reexpands towards the horizon $H^{-1}$. 

However, there is yet another possibility. Quantum mechanically, the loop can follow the first trajectory, penetrate the potential barrier from $R_{1}$ to $R_{2}$ and start expanding. The tunneling probability for this process can be estimated in the WKB approximation as:
\beq
P\sim e^{-B}~,
\eeq
where $B=2\int|p_{R}|dR$ integrating over the two turning points $R_{1}$ to $R_{2}$.

We can compute the tunneling probability explicitly with the help of (\ref{mom}) and (\ref{rdotv}):
\beq
P\sim\exp\left(-4\pi\mu{\int_{R_{1}}}^{R_{2}}\frac{dR}{f(R)}\sqrt{R^{2}f(R)-\epsilon^{2}}\right)~.
\eeq
In this paper, we are mainly interested in the spontaneous nucleation of loops, meaning loops that tunnel from zero size and have zero energy. To calculate the tunneling rate for such process we take the limit $\epsilon\rightarrow0$. The turning points in this case become, $R_{1}=0$ and $R_{2}=H^{-1}$, while the nucleation rate acquires the finite, non zero value:
\beq
P\sim\exp\left(-\frac{4\pi\mu}{H^{2}}\right)~.
\label{rate}
\eeq
This process describes the spontaneous nucleation of a loop of string of de Sitter radius with a nucleation rate given by (\ref{rate}). We note that our result is valid as long as the semiclassical approximation holds, that is, $\mu/H^{2}\gg1$ and as long as the string core thickness is much smaller than the de Sitter radius so we can use the thin wall approximation. Similar expressions for the nucleation rates of different objects in de Sitter space have been obtained for domain walls, monopoles \cite{Basu:1991ig} and black holes \cite{Bousso:1998na}.

It is well known that quantum tunneling processes can be effectively described using instantons \cite{Coleman:1985rnk,Coleman:1980aw}. The present scenario can similarly be examined from this perspective. In particular, the action derived previously indicates that the appropriate instanton representing the nucleation of a string in de Sitter space corresponds to the area of a two-dimensional Euclidean sphere representing the worldsheet of the string wrapping the equator of the four-dimensional sphere that characterizes Euclidean de Sitter space. It can be readily demonstrated that the equations of motion for the Nambu-Goto action in Euclidean space yield a solution consistent with this description. 

%Take the Euclidean static patch of deSitter:
%\beq
%ds_{E}^{2}=f(r)d\tau^{2}+\frac{dr^{2}}{f(r)}+r^{2}d\Omega^{2}
%\eeq
%A change of coordinates $Hr=\cos\omega$ brings the metric to the form:
%\beq
%ds_{E}^{2}=\frac{1}{H^{2}}\left[d\omega^{2}+\sin^{2}\omega d\tau^{2}+\cos^{2}\omega \left(d\theta^{2}+\sin^{2}\theta %d\phi^{2}\right)\right]
%\eeq
%This corresponds to the metric of a 2-sphere of radius $H^{-1}$ and a 2-sphere of radius $H^{-1}\cos\omega$. We can %associate the latter with the worldsheet of the string. In that case the Nambu-Goto action becomes:
%S_{NG}^{(E)}= -\frac{4\pi\mu}{H^{2}} \cos^{2}\omega
%\eeq
%which is minimized for $\omega=0$.Thus, the worldsheet of the string recasts to a 2-sphere of deSitter radius:
%\beq
%ds_{E}^{2}=\frac{1}{H^{2}}\left(d\theta^{2}+\sin^{2}\theta d\phi^{2}\right)
%\eeq

This simplified form of the instanton will serve as a basis for identifying new deformed solutions once we relax some of the  assumptions employed to derive this straightforward solution.
In our analysis, we aim to incorporate several factors that have been omitted thus far, which stem 
from the simplified approach used when describing our global string with the Nambu-Goto term.

In particular, it is clear that in certain regions of the parameter space, we cannot treat the string as thin, as its thickness, in a field theory description, becomes comparable to the other relevant scale in the problem, specifically the size of the horizon. Hence, at least for these cases we need to go beyond the simple description given here\footnote{This effect has been considered in detail for domain walls in \cite{Basu:1992ue}.  However, the string case was only discussed in the limiting scenario where the string thickness equals the de Sitter horizon. In this work, we will consider the full range of possible values of these parameters.}.
Furthermore, our instantons will also incorporate the effects of the global string's coupling to the Goldstone mode in relation to the instanton solutions and their corresponding action.

Lastly, we seek to understand the gravitational backreaction on the instantons associated with the global string. While this has been examined in the context of a thin local string in \cite{Basu:1991ig}, it is well known that the coupling of global strings to the massless mode induces significant gravitational effects on the global string metric \cite{global-string-singularities}. Therefore, we anticipate that this coupling will also play an important role in our solutions.

To account for all these effects in our analysis, we transition in the following section to a smooth description of the global string within the framework of a scalar field theory coupled to gravity. This approach requires numerical integration of the equations of motion, but wherever feasible, we will compare our solutions to the simplified ones presented here to better understand the origin of any deviations.

\section{Quantum Nucleation of Field Theory Global Strings}

To derive the instanton solutions representing the nucleation of global strings in de Sitter 
space within field theory, we must first define the specific model under consideration. The Euclidean action for the model we consider is expressed as follows:
\beq
S_{E}=\int\sqrt{g}~d^{4}x\left(-\frac{R}{16\pi G}+\rho_{v}\right)+S_E^{s}~,
\label{G-action}
\eeq
where we have included the gravitational Einstein-Hilbert term, a cosmological term and the string action given in terms
of the complex scalar field, $\phi$, namely:
\beq
S_E^{s}=\int{ d^{4}x ~\sqrt{g} ~\Big[ \p^{\mu}\phi~\p_{\mu}\phi^{\ast}+V(\phi)\Big]}~,
\label{string action}
\eeq
and where the vacuum energy $\rho_v$ and string potential can be respectively given by,
\beq
\rho_{v}=\frac{3H^{2}}{8\pi G}~~~~;~~~~V(\phi)=\lambda\left(|\phi|^{2}-\eta^{2}\right)^{2}~.
\eeq

Our model posses three parameters, the de Sitter energy density $\rho_{v}$ (or alternatively $H$) 
and the parameters that specify the string potential, $\lambda$ and $\eta$. As we will show in the following, these parameters introduce three distinct scales, namely the de Sitter radius, the string core thickness and the induced spacetime horizon due to the gravitational interaction of the defect. We will comment more about each scale and their effect in our solutions as we proceed forward with our analysis.  

The appropriate ansatz for the Euclidean metric of the instantons we are looking for should match the symmetries of the 
thin wall solution described. In other words our ansatz for the geometry should reduce itself to the simple instanton presented earlier in the thin wall limit. Therefore we are led to the following form of the metric,
\beq
ds_{E}^2=dr^{2}+a^{2}(r)\left(d\tau^{2}+\sin^{2}\tau~d\chi^{2}\right)+b(r)^{2}d\theta^{2}~,
\label{E-metric}
\eeq
where $a(r)$ and $b(r)$ are scale factors and we take $0\le\tau\le\pi$ and $0\le\chi,\theta\le2\pi$. We also parametrize the complex scalar field as:
\beq
\phi=\varphi(r)e^{i\theta}
\eeq
where $\theta$ is the angle describing the $S^{1}$ part of the geometry in Eq. (\ref{E-metric}). Note that
with this ansatz the string worldsheet is parametrized by the ($\tau$, $\chi$) coordinates and describes a $S_2$ sphere, as in the thin wall limit, while the
$(r,\theta)$ part of the geometry represents the $2d$ manifold perpendicular to the string. Furthermore the scalar field ansatz is consistent with the presence of a vortex configuration around the region where $b(r) \rightarrow 0$ and its particular form assumes that the vortex winding number is one\footnote{Here we will focus our discussion in the unit winding number. There is in principle no impediment to find similar solutions to higher winding numbers.}. This is the position of the center of our global string.

The Euclidean action can be calculated by inserting the ansantz (\ref{E-metric}) in (\ref{G-action}). It is
straightforward to show that action becomes,
\beq
S^G_{E}=-\frac{\pi}{ G}\int dr\left[b\left({a^{\prime}}^{2}+1\right)+2aa^{\prime}b^{\prime}-3H^{2}a^{2}b\right]+S_{b}~,
\label{action}
\eeq
where we carried out the angular integration and also derived a boundary term as a result of the integration by parts:
\beq
S_{b}=\frac{\pi}{G}\left[\frac{d}{dr}\left(a^{2}b\right)\right]~.
\eeq
At this point we will not specify the integration bounds for the radial component $r$, but in the proceeding sections it will be explicitly defined as the string core up to the deformed de Sitter horizon.  We proceed by evaluating the string Euclidean action in the same manner:
\beq
S_E^{s}=8\pi^{2}\int dr a^{2}b\left[{\varphi^{\prime}}^{2}+\frac{\varphi^{2}}{b^{2}}+V(\varphi)\right]~.
\eeq
The action comprises of three terms, namely the kinetic energy of the massive (radial) mode of the scalar field, its potential and the gradient, which appears due to the coupling to the massless Goldstone. We note that the latter contributes logarithmically to the energy momentum tensor necessitating a physical cutoff for the computation of the string tension in asymptotically flat space \cite{Vilenkin:2000jqa}.

The Euclidean equations of motion are derived by the varying the above expressions with respect to $a(r),~b(r)$ and $\varphi(r)$. The result of this procedure yields :
\beq
2aa^{\prime\prime}+{a^{\prime}}^{2}-1+3H^{2}a^{2}=-8\pi Ga^{2}\left({\varphi^{\prime}}^{2}-\frac{\varphi^{2}}{b^{2}}+V(\varphi)\right)~,
\label{eqofm}
\eeq

\beq
ab^{\prime\prime}+ba^{\prime\prime}+a^{\prime}b^{\prime}+3H^{2}ab=-8\pi Gab\left({\varphi^{\prime}}^{2}+\frac{\varphi^{2}}{b^{2}}+V(\varphi)\right)~,
\label{eqofm1}
\eeq
and
%Varying with respect to $\phi(r)$ yields:
\beq
\varphi^{\prime\prime}+\left(\frac{b^{\prime}}{b}+\frac{2a^{\prime}}{a}\right)\varphi^{\prime}=\frac{\varphi}{b^{2}}+\frac{1}{2}\left(\frac{\p V}{\p\varphi}\right)~.\\
\eeq
\textbf{}

Finally, the Hamiltonian constraint equation reads:
\beq
{a^{\prime}}^2+2aa^{\prime}\frac{b^{\prime}}{b}-1+3H^{2}a^{2}=8\pi Ga^{2}\left({\varphi^{\prime}}^{2}-\frac{\varphi^{2}}{b^{2}}-V(\varphi)\right)
\eeq
Only three of the above equations are independent, as the constraint is essentially the integration of (\ref{eqofm}) and (\ref{eqofm1}).

%\beq
%{a^{\prime}}^{2}=1-\Lambda_{eff}a^{2}-\frac{8\pi G}{a}\int a^{2}da\left({\phi^{\prime}}^{2}-\frac{\phi^{2}}{b^{2}}+\lambda\phi^{4}-2\lambda\eta^{2}\phi^{2}\right)
%\eeq

\subsection{Boundary Conditions}

A key step in solving the equations given above is to impose appropriate boundary conditions that are consistent with the instantons we are looking for.

Let us first consider the $r=0$ region with the following conditions,
\beq
a(0)=a_{0}~,~~b(0)=0~,~~\phi(0)=0~.
\eeq
The fact that at $r=0$ the function $b(r)$ vanishes and the field climbs to the
top of the potential are clearly signaling that this region corresponds to the core of the global string. Additionally, the interpretation of the parameter $a_0$ becomes clear: it defines the size of the sphere representing the Euclidean worldsheet of the string span by the $\tau, \chi$ coordinates.

On the other end of the range of the radial coordinate $r$, we impose
\beq
a(r_{\ast})=0~,~~b(r_{\ast})=b_{\ast}~,~~\varphi(r_{\ast})=\varphi_{\ast}~.
\label{bc1}
\eeq
This choice is motivated by the case where the gravitational backreaction of the string 
is negligible. In this limit the manifold should be approximately given by de Sitter space. Indeed, there 
exists a simple vacuum solution with the same symmetry, which is essentially pure de Sitter 
space. In Appendix \ref{deSitter-appendix}, we provide a detailed discussion of this anisotropic formulation 
of de Sitter space and outline the corresponding boundary conditions for comparison with the 
current solution.

Finally we note that in order to ensure the absence of any singular behaviour of the metric we also need to impose
\beq
a^{\prime}(r_{\ast})=\pm1~,~~b^{\prime}(r_{\ast})=0~,~~a^{\prime\prime}(r_{\ast}=0) ~~,
\label{bc2}
\eeq
as well as,
\beq
\label{bc3}
b^{\prime}(0)=\pm1~,~~a^{\prime}(0)=0~,~~b^{\prime\prime}(0)=0~~.
\eeq

Furthermore, regularity of the equation of motion for $\phi$ imposes the condition:
\beq
\varphi^{\prime}(r_{\ast})=0~.
\label{phidot at horizon}
\eeq
Given these conditions, the regularity of the metric is guaranteed. In fact, one can
show that the region $r=r_*$ is a horizon and the metric can be extended beyond this
region in a smooth way given these boundary conditions. We will comment on this fact later on
in the paper.

\section{Instanton Solutions}

The next step is to solve the above equations for different values of the parameters $H$, $\lambda$ and $\eta$. In particular we plan to solve the following system of equations for the scale factors $a$, $b$ and the field profile $\varphi$:

\beq
2aa^{\prime\prime}+{a^{\prime}}^{2}-1+3H^{2}a^{2}=-8\pi Ga^{2}\left({\varphi^{\prime}}^{2}-\frac{\varphi^{2}}{b^{2}}+\lambda\left(\varphi^{2}-\eta^{2}\right)^{2}\right)~,
\label{3x3system1}
\eeq

\beq
\varphi^{\prime\prime}+\left(\frac{b^{\prime}}{b}+\frac{2a^{\prime}}{a}\right)\varphi^{\prime}=\frac{\varphi}{b^{2}}+\lambda\varphi\left(\varphi^{2}-\eta^{2}\right)~,
\label{3x3system2}
\eeq

\beq
{a^{\prime}}^2+2aa^{\prime}\frac{b^{\prime}}{b}-1+3H^{2}a^{2}=8\pi Ga^{2}\left({\varphi^{\prime}}^{2}-\frac{\varphi^{2}}{b^{2}}-\lambda\left(\varphi^{2}-\eta^{2}\right)^{2}\right) ~.\\
\label{3x3system3}
\eeq

\vspace{0.4cm}

For future reference, we will define the following dimensionless quantities that are related to our 
initial input parameters and possess straightforward physical interpretations. The classification of 
solutions will be based on the values of these parameters.

The first one is the squared of the ratio between the de Sitter horizon $H^{-1}$ and the string core thickness $\delta=(\lambda\eta^{2})^{-1/2}$, 
\beq
C=\frac{\lambda\eta^{2}}{H^{2}}~.
\eeq

The other interesting quantity is a measure of the string's gravitational backreaction to the geometry,
and is given by,
\beq
D=8\pi G\eta^{2}~.
\eeq

\subsection{Nucleation in a flat background}

As a first step, we study solutions of our system of equations for global strings in a flat background spacetime \footnote{
We begin by studying these solutions to highlight the significant new gravitational effects that arise in the global string configurations. Additionally, as we will demonstrate later, these solutions can be interpreted as instantons in the context of quantum cosmology.}. Accordingly, we set $H = 0$. For the sake of convenience we will rescale our variables in terms of the characteristic length scale, $\delta$, and introduce the dimensionless description of the scalar field in terms of $y=\varphi/\eta$.  This reduces the equations of motion to a system dependent on a single parameter ($D$):
\beq
2aa^{\prime\prime}+{a^{\prime}}^{2}-1=-Da^{2}\left({y^{\prime}}^{2}-\frac{y^{2}}{b^{2}}+\left(y^{2}-1\right)^{2}\right)
\eeq
\beq
y^{\prime\prime}+\left(\frac{b^{\prime}}{b}+\frac{2a^{\prime}}{a}\right)y^{\prime}=\frac{y}{b^{2}}+2y\left(y^{2}-1\right)
\eeq
\beq
{a^{\prime}}^2+2aa^{\prime}\frac{b^{\prime}}{b}-1=Da^{2}\left({y^{\prime}}^{2}-\frac{y^{2}}{b^{2}}-\left(y^{2}-1\right)^{2}\right)
\eeq
where:
\beq
a\rightarrow \frac{a}{\delta}~,~~b\rightarrow\frac{b}{\delta}~,~~r\rightarrow\frac{r}{\delta}
\eeq

It is important to note, that the above system does not differ qualitatively from the de Sitter case we will present later on.  The backreaction of a global string to the geometry induces the presence of a horizon, even in the absence of a background positive cosmological constant. As such, we can proceed with numerically solving the system with the same boundary conditions as in the de Sitter case for the string core and horizon.

This effect has previously been identified in the Lorentzian formulation of spacetime 
surrounding a global string. In that context, it has been shown that the energy-momentum 
tensor associated with the winding mode around the string prevents the existence of a 
smooth, static, singularity-free spacetime \cite{global-string-singularities}. This issue 
is resolved by allowing the induced metric on the global string to become time-dependent, 
specifically adopting a de Sitter-like form for the string worldsheet \cite{Gregory:1996dd}. 
Our Euclidean configuration can be understood as the analytic continuation of such a solution, 
where the de Sitter-like configuration is mapped into a sphere representing the string’s worldsheet 
in Euclidean space.

Although the mathematical structure of the solution is the same, our interpretation is 
quite different from the Lorentzian case. As we will discuss in detail, the flat-space 
solution can be understood as an instanton in quantum cosmology representing the nucleation 
of a universe from nothing with a global string—analogous to similar studies involving domain walls \cite{Vilenkin:1983ykn, Ipser:1983db, Garriga:1999bq,Blanco-Pillado:2019tdf,Fanaras:2023acz}.

%Before we proceed with the solution we note some qualitative features of the geometry, depending on the values of the parameter $D$. First, in the marginal case of $D\rightarrow0$ we retrieve the flat space metric with solution:
%\beq
%a=a_{0}-r~,~~b=const
%\label{flatsol}
%\eeq

%In that case, there is no horizon induced by the defect, or to be more %precise, the horizon is displaced all the way to infinity. The opposite %occurs for values of $D\sim1$. In particular,there exists a homogeneous %configuration for which the horizon is displaced near the core of the %string. 

Before presenting the numerical solutions, we note that the current setup contains 
only two relevant length scales: the thickness of the string and the horizon 
distance induced by the string. It is therefore worthwhile to examine the behavior 
of the solutions as a function of these two scales.

In the limit where $D\rightarrow 0$, the horizon is pushed to infinity. This case 
can be interpreted as the regime where gravity is effectively decoupled, with the scalar
field profile being localized in a small region around the region $r\approx 0$. The horizon, being far removed, 
does not significantly influence the solution in this regime.

Conversely, in the limit where $D\sim \mathcal{O}(1)$, the horizon distance becomes comparable
to the string thickness. In this region of the parameter space, the geometry is 
strongly distorted by the string, and the field profile approaches the top of 
the potential across the entire region. This leads to a solution of the Hawking-Moss
type \cite{Hawking:1981fz}, corresponding to the regime of topological inflation, where the 
effective cosmological constant associated with the potential energy induces a 
horizon distance on the order of the string thickness \cite{Vilenkin:1994pv,Linde:1994wt}. In this case, the 
field configuration becomes nearly homogeneous. We will discuss in detail and derive the condition for the appearance of the homogeneous instanton in Sec. 4.3 .

%In this case, following the same steps as in subection 2.3 we write equations of motion for the scale factors for $y \approx 0$:
% \beq
%2aa^{\prime\prime}+{a^{\prime}}^{2}-1+Da^{2}=0
%\eeq
%and
%\beq
%{a^{\prime}}^2+2aa^{\prime}\frac{b^{\prime}}{b}-1+Da^{2}=0
%\eeq
 %which can be solved to:
 %\beq
%a=\frac{\cos\tilde{r}}{\sqrt{\frac{D}{3}}}~,~~b=\frac{\sin\tilde{r}}{\sqrt{\frac{D}{3}}}
%\label{homflat}
%\eeq
%where $\tilde{r}=r\sqrt{\frac{D}{3}}$ and its range is $\tilde{r}\in\left[0,\pi/2\right]$. It is easy to see that one need not repeat all the steps to the derivation, but rather we substitute $1 + CD/3$ with $D/3$ in the final result. Doing so yields the condition:
 %\beq
 %D=3/2
 %\eeq
 %This can be interpreted as the flat space limit of (1.33) in the main draft.

\subsubsection{The solutions}

\begin{figure}[t]
  \begin{subfigure}[t]{0.475\textwidth}
    \includegraphics[scale=0.6]{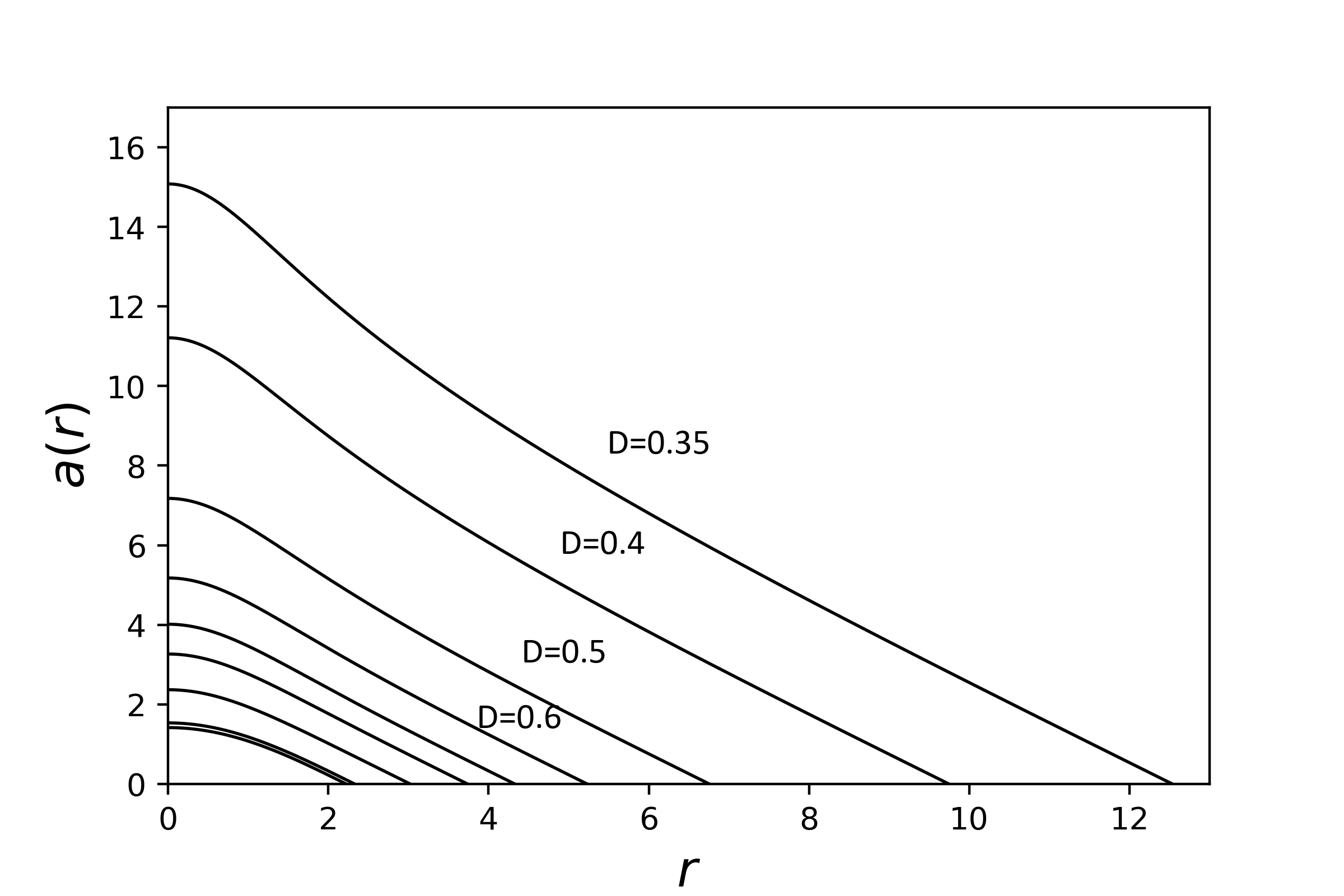}
    \caption{Profiles $a(r)$ for different values of $D$ }
      \end{subfigure}\hfill
  \begin{subfigure}[t]{0.475\textwidth}
    \includegraphics[scale=0.6]{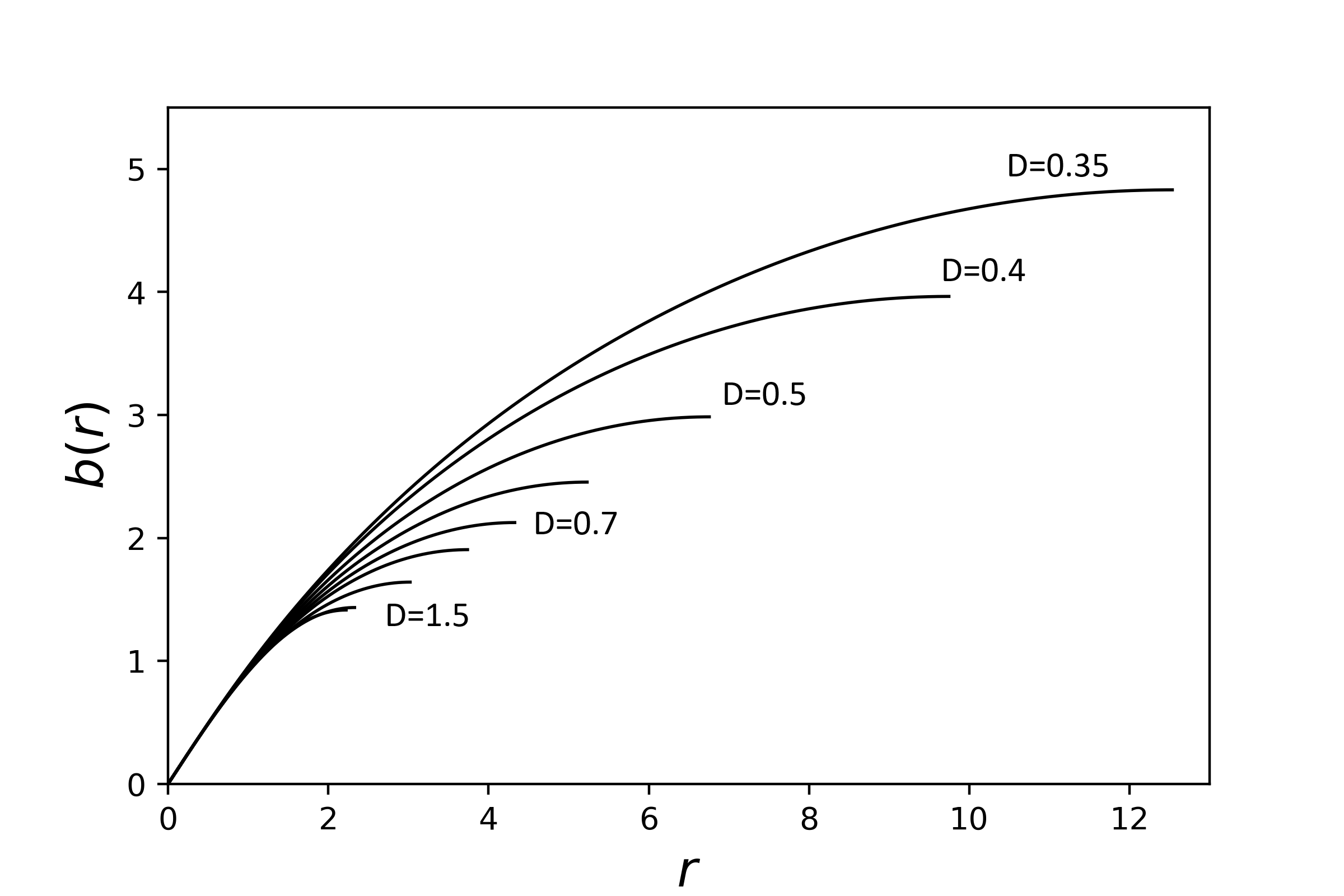}
    \caption{Profiles $b(r)$ for different values of $D$ }
      \end{subfigure}
  \begin{subfigure}[t]{0.475\textwidth}
    \includegraphics[scale=0.6]{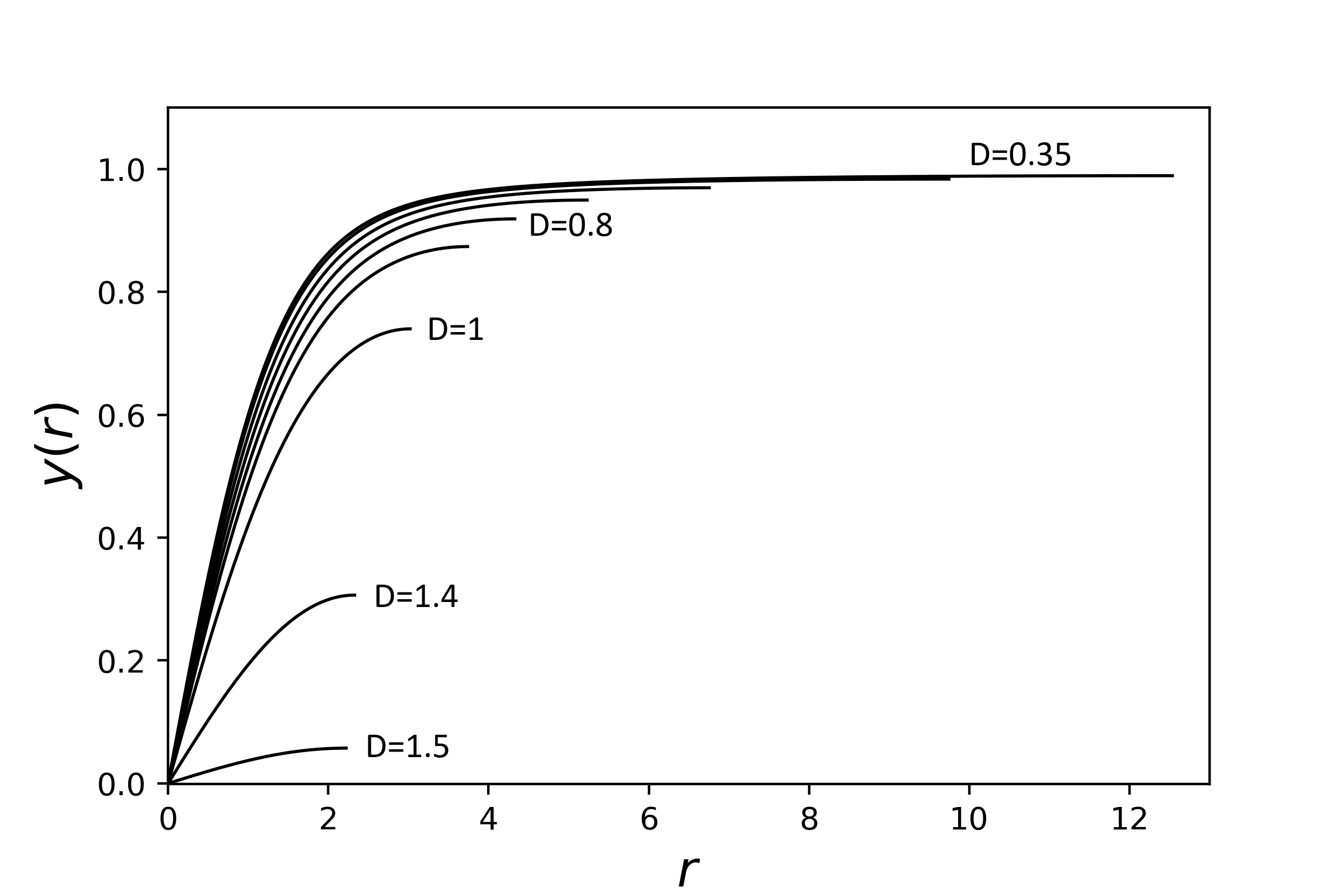}
    \caption{Profiles $y(r)$ for different values of $D$ }
      \end{subfigure}       
  \caption{Plots of the profiles $a(r)$, $b(r)$ and $y(r)$ for varying $D$. We start with values of $D\rightarrow3/2$ and decrease till $D\sim0.1$. In the former case the geometry is equivalent to de Sitter and $y\approx 0$. As $D$ decreases the horizon is pushed to larger and larger distances and the scalar field profile resembles the one in a fixed flat background.} 
  \label{fig1flat}
\end{figure}

\begin{figure}[b]
  \begin{subfigure}[t]{0.475\textwidth}
    \includegraphics[scale=0.6]{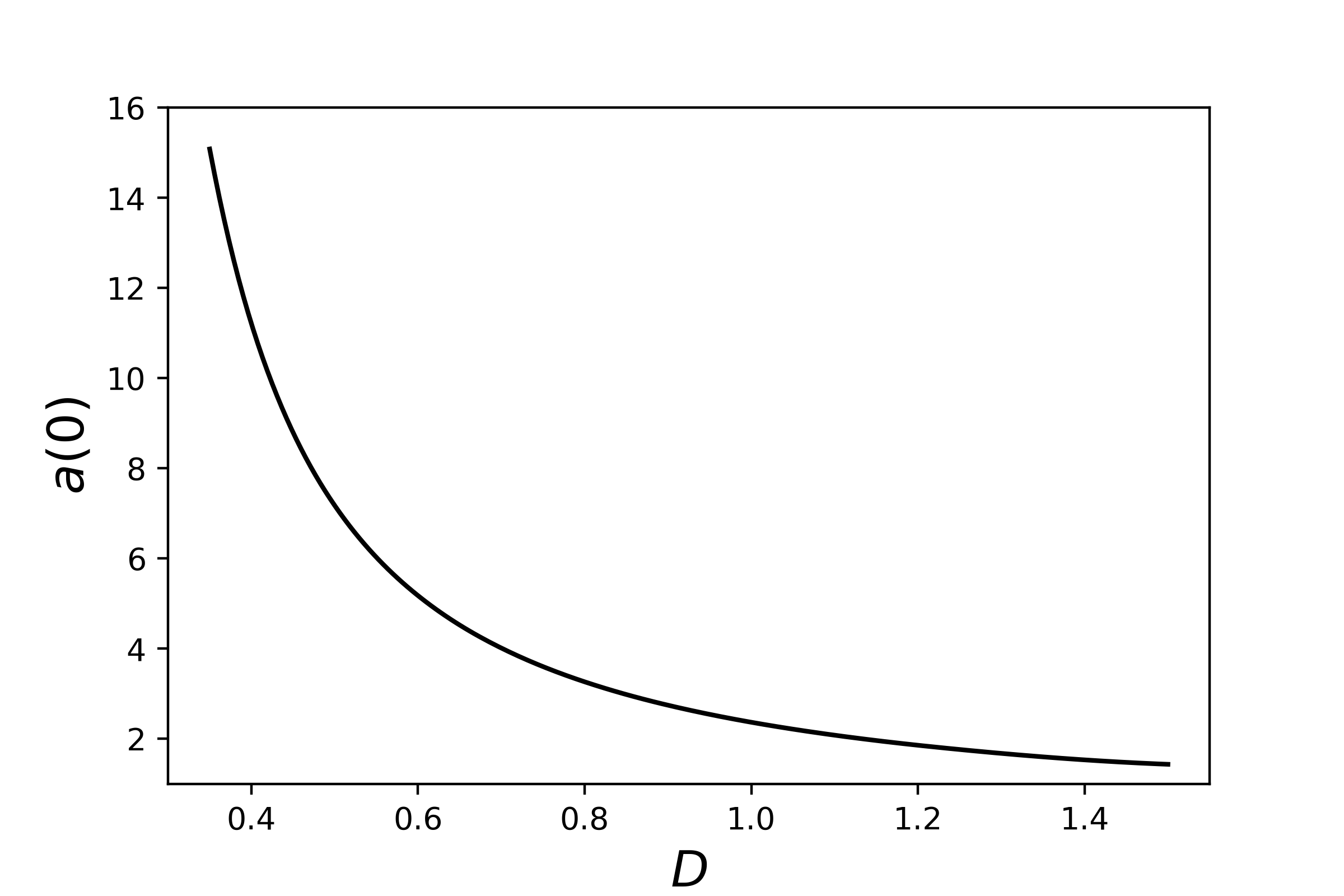}
    \caption{Plot of $a(0)$ vs $D$ }
      \end{subfigure}\hfill
  \begin{subfigure}[t]{0.475\textwidth}
    \includegraphics[scale=0.6]{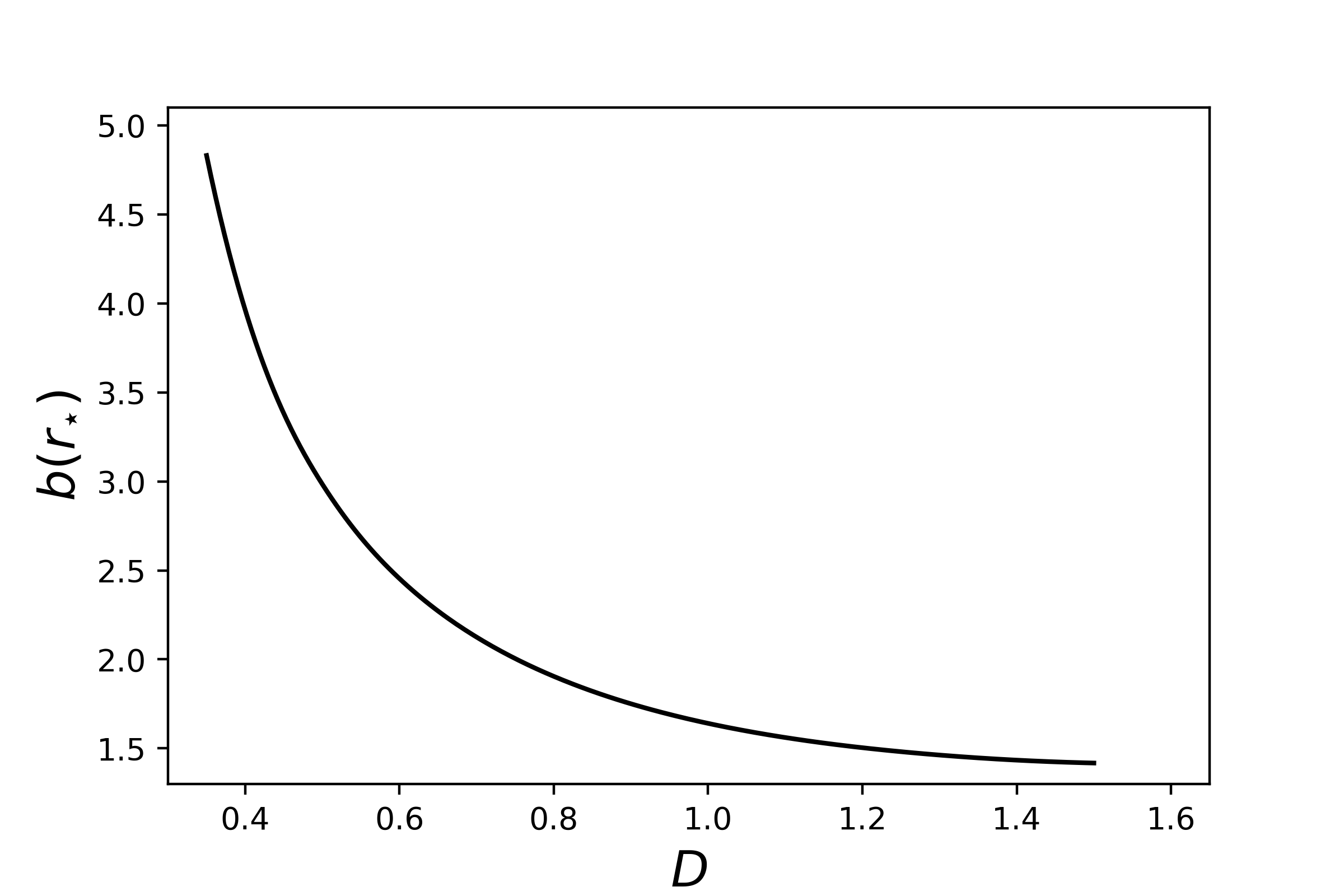}
    \caption{Plot of $b(r^{\ast})$ vs $D$ }
      \end{subfigure}
  \begin{subfigure}[t]{0.5\textwidth}
    \includegraphics[scale=0.58]{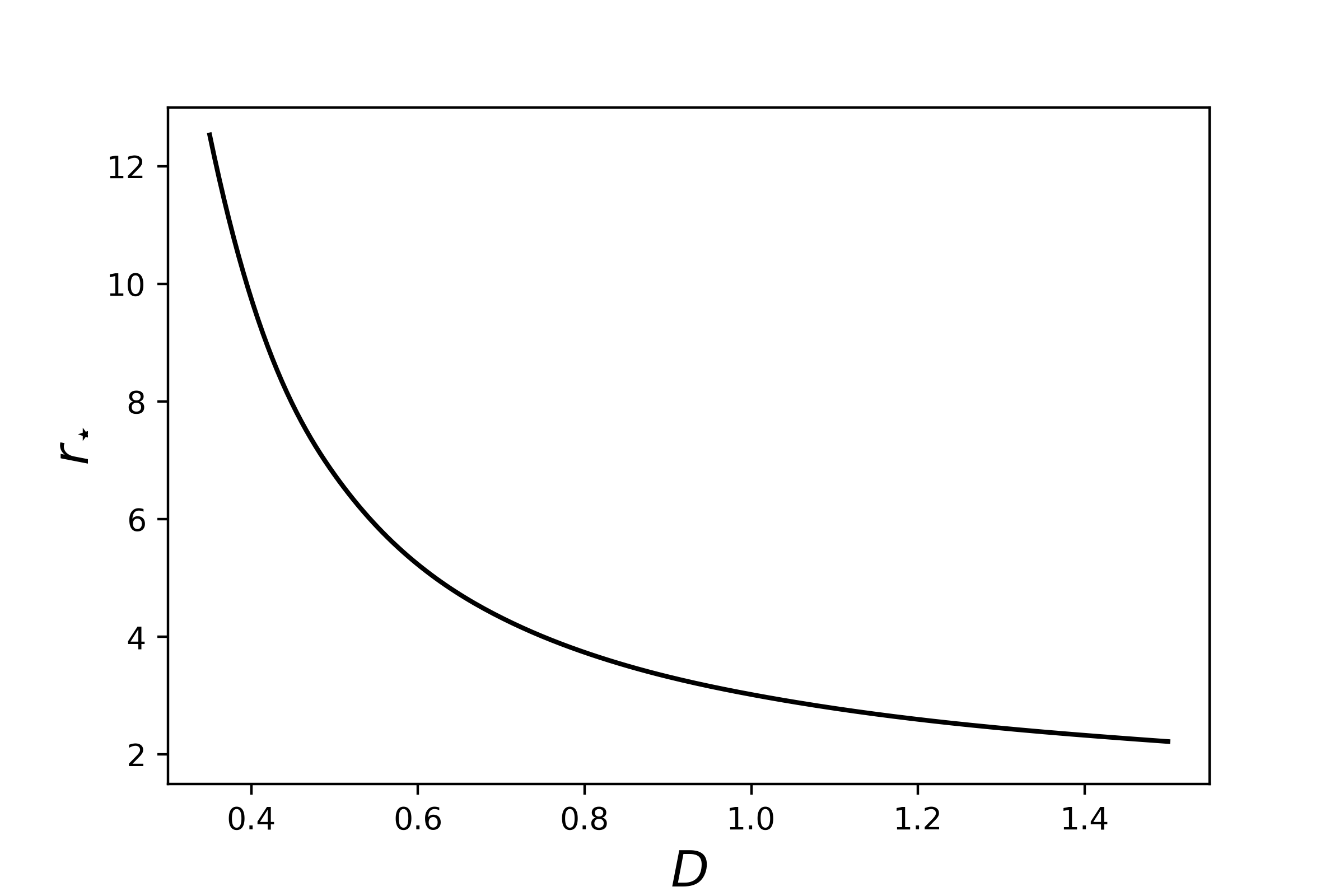}
    \caption{Plot of $r_{\ast}$ vs $D$ }
      \end{subfigure}
       
  \caption{Plots of $a_{0},~b_{\ast}$ and $r_{\ast}$ in terms of $D$. As $D\rightarrow\frac{3}{2}$ we approach the Hawking-Moss configuration. For $D\sim0.1$ the scale factors and the radial displacement grow since we approach the flat space solution.  } 
  \label{fig2flat}
\end{figure}

We begin by exploring the behavior of the geometry and the scalar field profile for different values of the coupling parameter $D$ \footnote{As previously mentioned, it is crucial to impose the correct boundary conditions to obtain the desired solutions. This necessitates the use of specific numerical methods to avoid potential numerical singularities. The detailed description of these methods is provided in Appendix \ref{numerics}, while in the following sections we just focus on the description of the solutions obtained using those techniques.}.The results are illustrated in Figs. \ref{fig1flat} and \ref{fig2flat}.

For minimal backreaction, the scalar field approaches the solution obtained by solving the equations of motion in a fixed flat background. Specifically, the core region has $y \approx 0$, but the field eventually reaches its vacuum expectation value, $y \approx 1$, for most of the regime where $r > 1$. However, as expected, the gravitational effect of the coupling between the vortex and the long-range Goldstone mode still induces a cosmological horizon, although it is displaced far from the string core whenever $D \ll 1$.

When the coupling parameter $D$ increases, the string distorts the geometry substantially rendering it effectively with the same structure as the anisotropic de Sitter discussed in Appendix \ref{deSitter-appendix}. In the marginal case where $D \approx 3/2$, the cosmological horizon is pushed inward, reaching the string core. As a result, the configuration becomes homogeneous, with $y \approx 0$ and a de Sitter background geometry.

Furthermore, we examine the geometrical distortions that the coupling $D$ induces. As expected, for $D\rightarrow0$ the range of $a$ and $b$ are much larger than the size of the defect. As the system couples more strongly to the background, and approaches its limiting case where $D = 3/2$, the geometry gradually compactifies with the radial displacement approaching the core of the vortex.  See Figs. \ref{fig2flat}.

\subsection{Nucleation in a de Sitter background}

Let us now consider the instanton solutions representing the nucleation of global strings 
in an external de Sitter background. First, we rescale the variables again, this time by the following transformations,
\beq
a\rightarrow Ha~,~~b\rightarrow Hb~,~~r\rightarrow Hr~,
\eeq
essentially measuring the lengths in terms of de Sitter units.
With this rescaling the system of equations takes the form:
\beq
2aa^{\prime\prime}+{a^{\prime}}^{2}-1+3a^{2}=-Da^{2}\left({y^{\prime}}^{2}-\frac{y^{2}}{b^{2}}+C\left(y^{2}-1\right)^{2}\right)~,
\label{eq1}
\eeq
\beq
y^{\prime\prime}+\left(\frac{b^{\prime}}{b}+\frac{2a^{\prime}}{a}\right)y^{\prime}=\frac{y}{b^{2}}+2Cy\left(y^{2}-1\right)~,
\label{scalar eom}
\eeq
\beq
{a^{\prime}}^2+2aa^{\prime}\frac{b^{\prime}}{b}-1+3a^{2}=Da^{2}\left({y^{\prime}}^{2}-\frac{y^{2}}{b^{2}}-C\left(y^{2}-1\right)^{2}\right)~,
\label{eq2}
\eeq

The interesting point about these equations is the simultaneous presence 
of three distinct length scales: the two previously discussed—the string thickness and the induced horizon scale—and the de Sitter horizon scale associated with a non-zero $H$. In the following section, we present some of the numerical solutions obtained, which provide deeper insight into the qualitative behavior of these instanton solutions across different regions of the parameter space.

\subsubsection{The solutions}

\begin{figure}[h!]
    \begin{subfigure}[t]{0.475\textwidth}
        \includegraphics[scale=0.6]{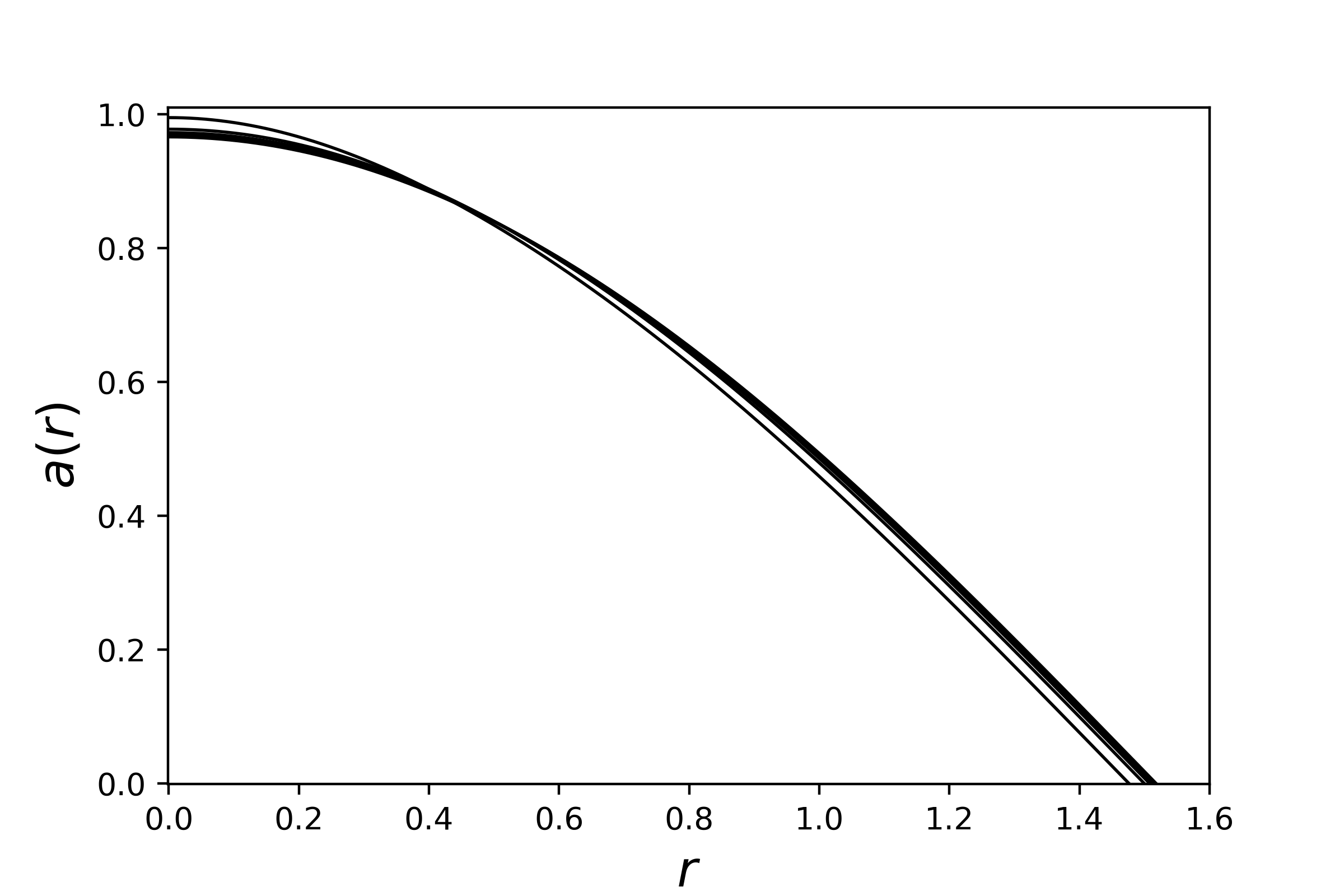}
        \caption{Plot of $a(r)$ vs $r$ for varying string thickness}
        \label{fig-1a}
    \end{subfigure}\hfill
    \begin{subfigure}[t]{0.45\textwidth}
        \includegraphics[scale=0.6]{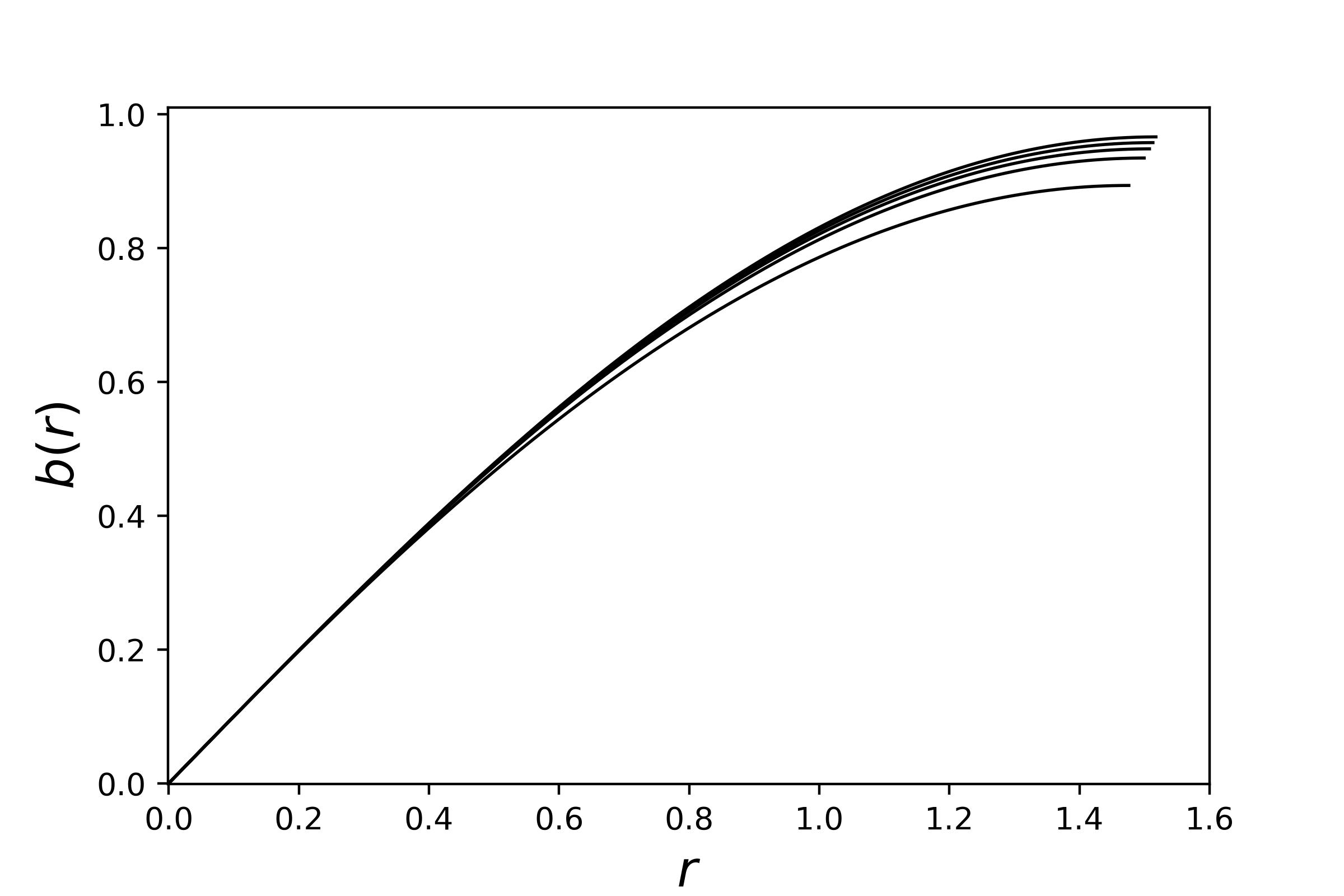}
        \caption{Plot of $b(r)$ vs $r$ for varying string thickness}
        \label{fig-1b}
    \end{subfigure}\hfill
    \begin{subfigure}[t]{0.6\textwidth}
        \includegraphics[scale=0.17]{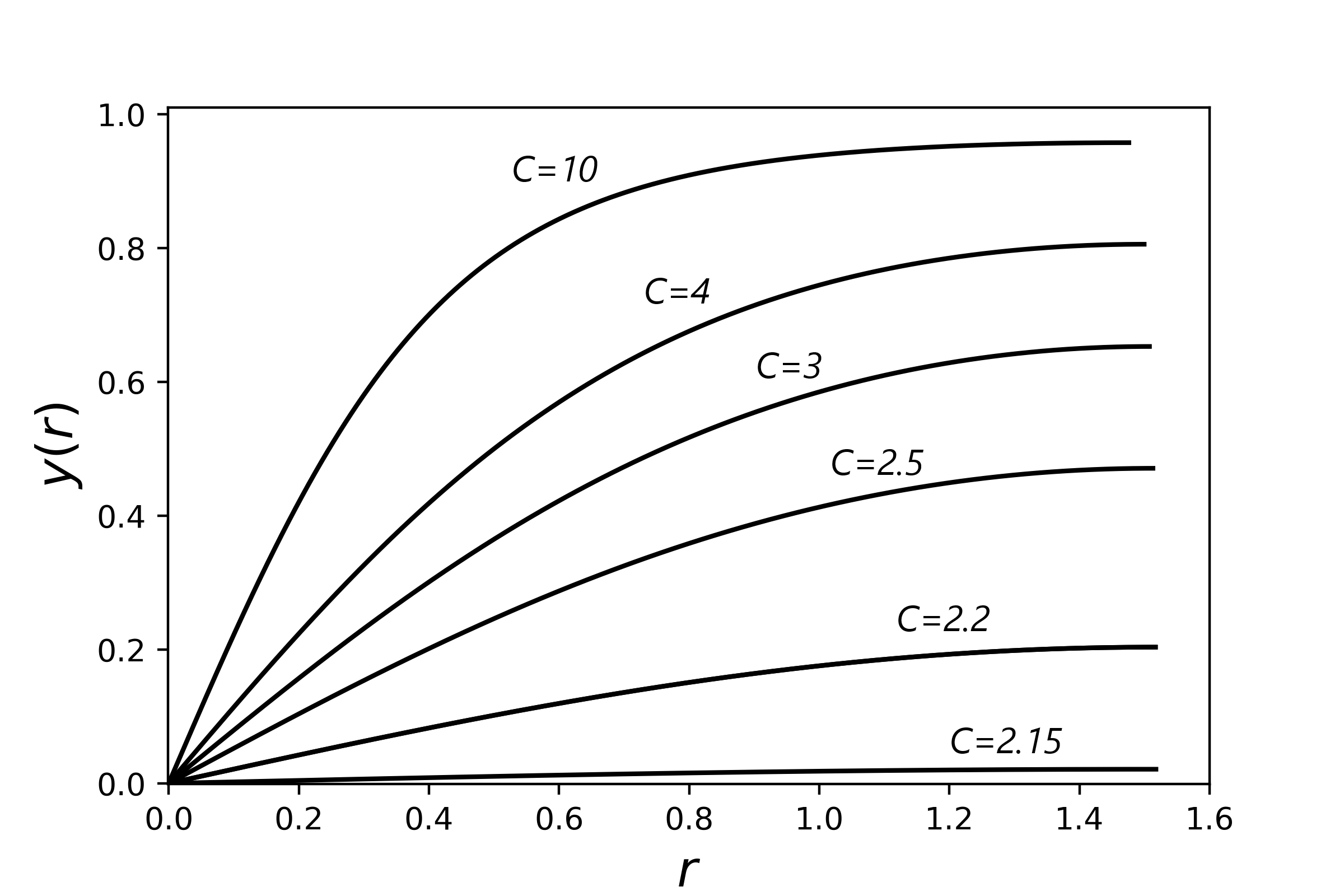}
        \caption{Plot of the field profile $y(r)$ vs $r$ for varying string thickness}
        \label{fig-1c}
    \end{subfigure}
    \caption{Plots of the profiles $a(r)$, $b(r)$ and $y(r)$ for fixed backreaction $D=0.1$ and varying thickness $C^{-1}$} 
    \label{fig1}
\end{figure}

In this section we obtain the solutions for different values of the parameters $C$ and $D$ and find once more the profiles $a(r)$, $b(r)$ and $y(r)$. An example is shown in Figs. \ref{fig1} where we kept the backreaction of the string to the geometry constant and relatively small, meaning we take $D=0.1$, and varied the thickness of the vortex core relative to the Hubble scale. As expected, when the backreaction is minimal, the thickness does not significantly deform the geometry. This is evident by the fact that $a_{0}$ and $b_{\ast}$ are relatively close to $1$, the value corresponding to de Sitter geometry.

\begin{figure}[b]
  \begin{subfigure}[t]{0.475\textwidth}
    \includegraphics[scale=0.6]{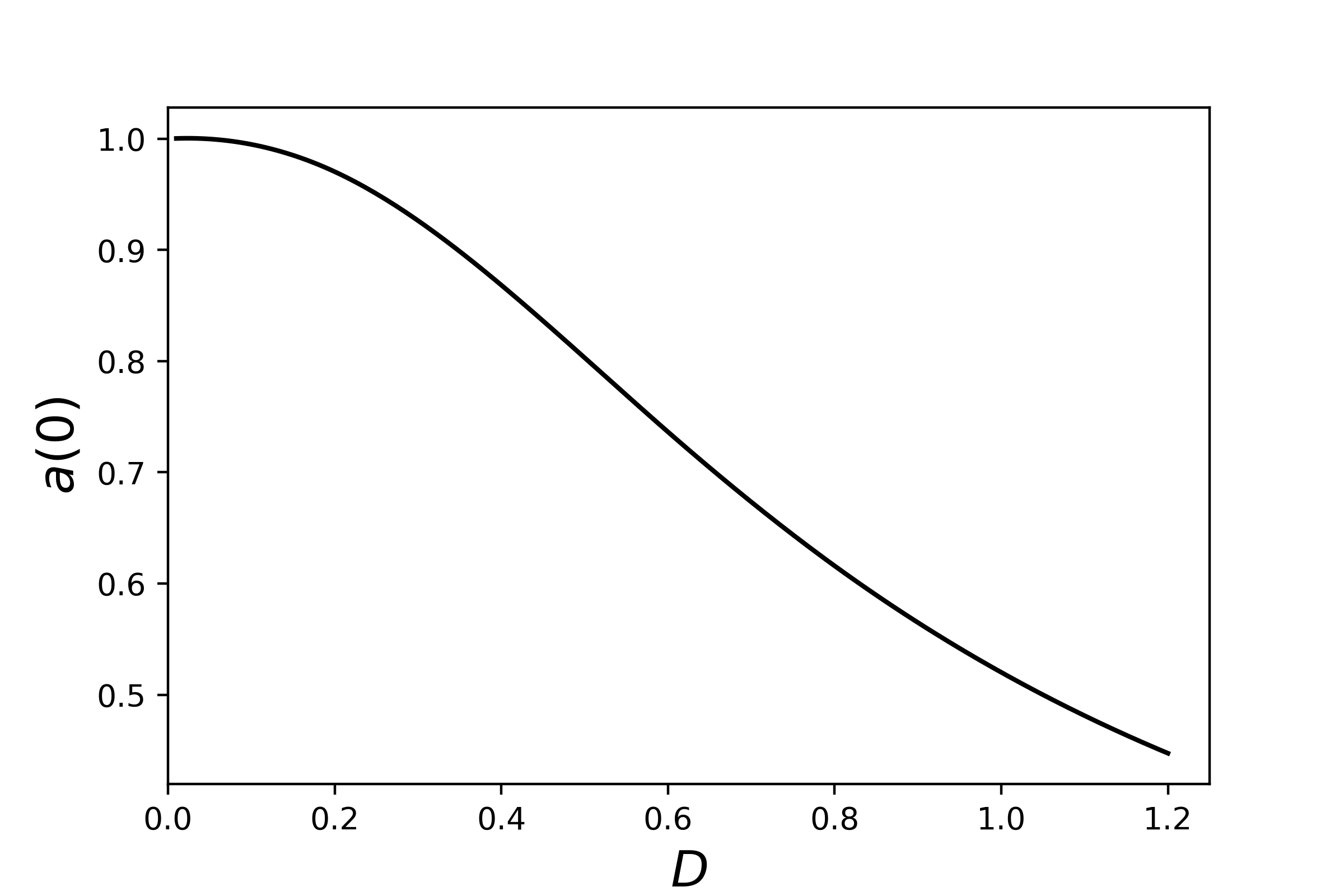}
    \caption{Plot of $a(0)$ vs $D$ }
    \label{fig-1a}
  \end{subfigure}\hfill
  \begin{subfigure}[t]{0.475\textwidth}
    \includegraphics[scale=0.6]{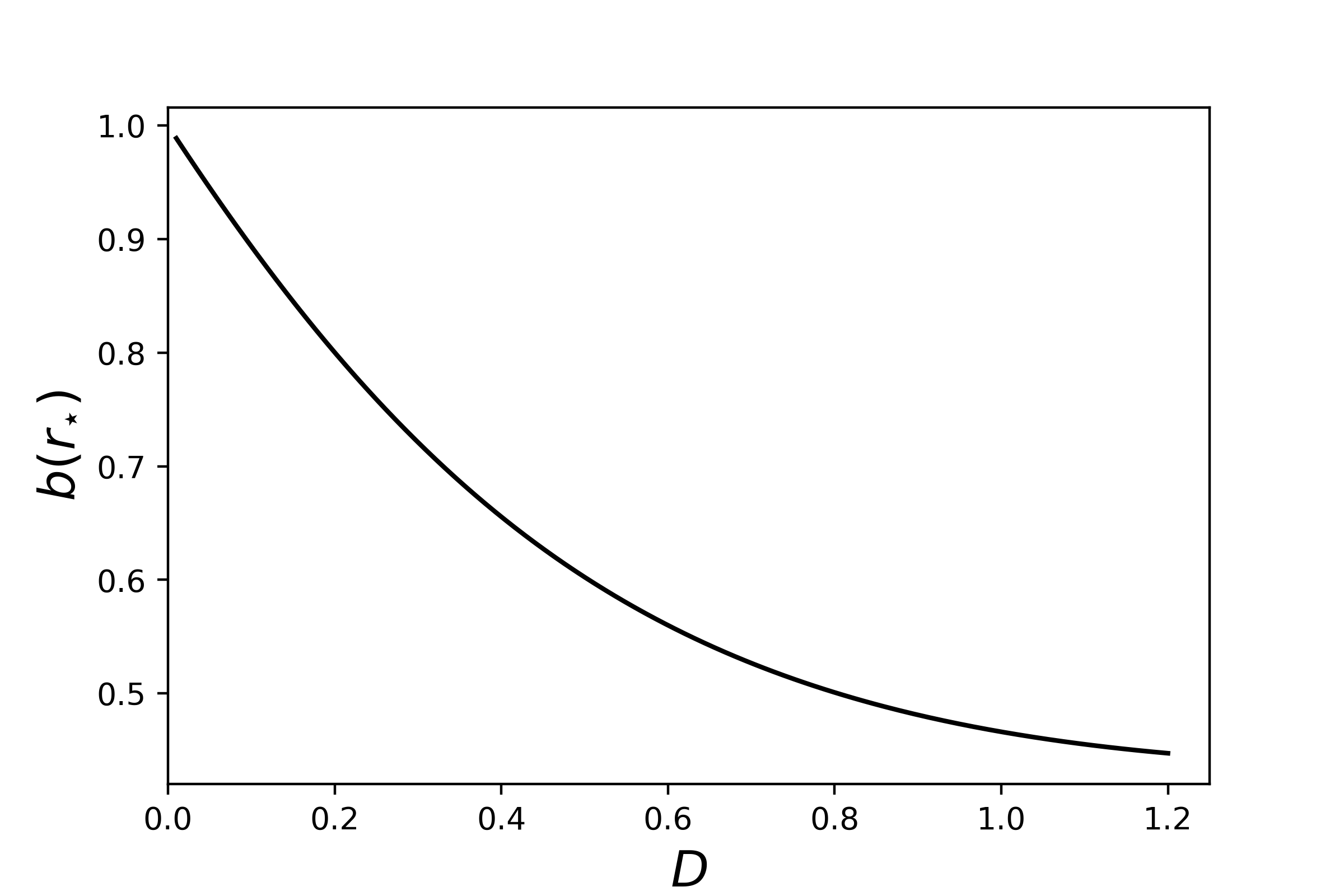}
    \caption{Plot of $b(r^{\ast})$ vs $D$ }
    \label{fig-1b}
  \end{subfigure}
  \begin{subfigure}[t]{0.475\textwidth}
    \includegraphics[scale=0.6]{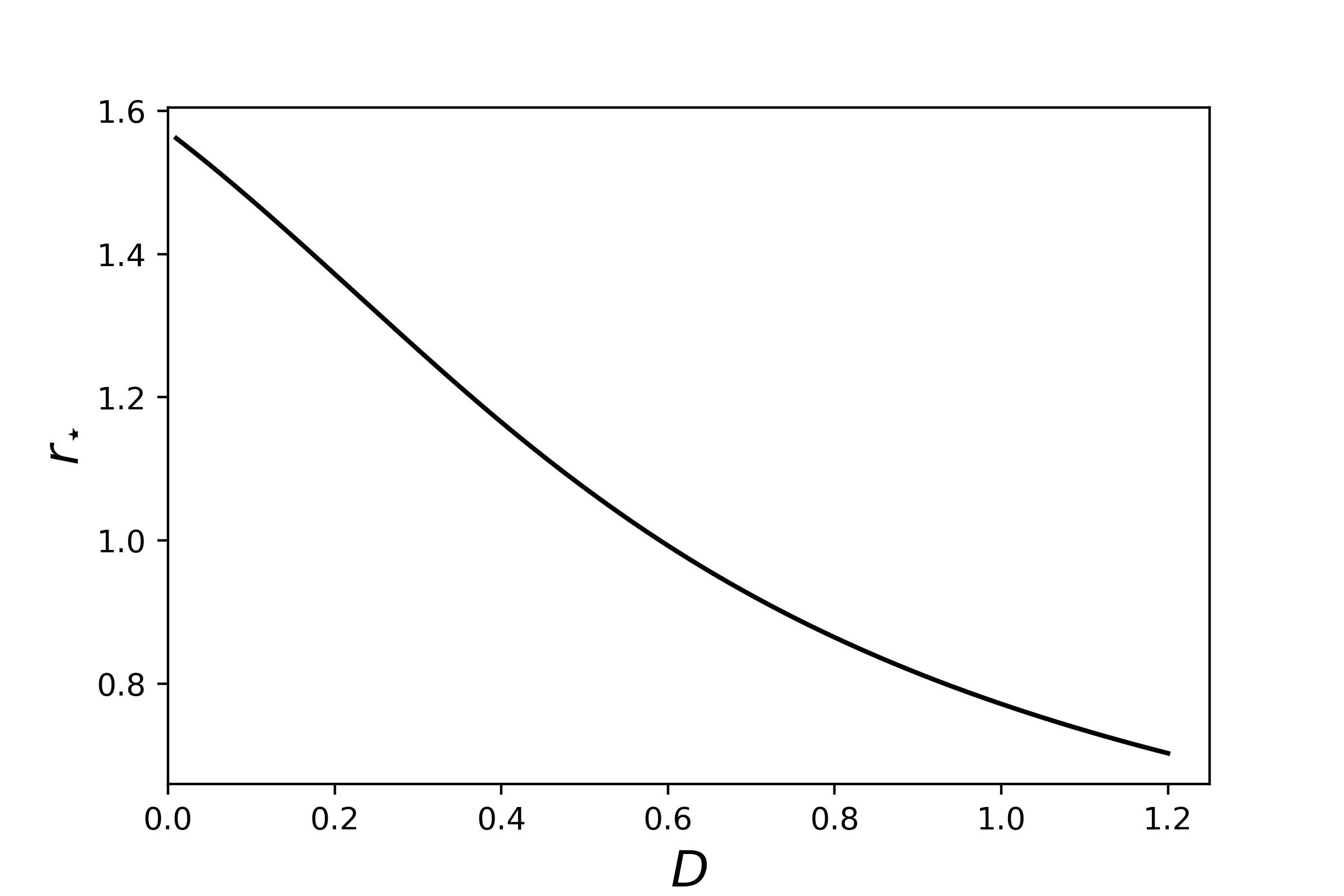}
    \caption{Plot of $r^{\ast}$ vs $D$ }
    \label{fig-1c}
  \end{subfigure}
  \caption{Plots of $a_{0},~b_{\ast},~r_{\ast}$ in terms of $D$. For no backreaction the geometry is pure de-Sitter while as $D$ increases the instanton gets substatially distorted.} 
  \label{fig2}
\end{figure}

However, the field profile is substantially altered for varying string thickness. When the core is thin the field approaches rapidly the vacuum value $y\rightarrow1$, while as the core thickens the field is closer to the maximum of the potential $y\approx0$. There exists a threshold for the thickness after which the configuration is homogeneous with $y(r)\approx0$. This occurs at a specific value of the parameters $C$ and $D$ as we will show in the next subsection.

In order to probe the effect of the string backreaction to the geometry we kept the thickness fixed at $C=10$ and varied the parameter $D$ from zero, which corresponds to pure de Sitter, to $D=1.2$ for which the homogeneous instanton appears. The distortion of the geometry is demonstrated in Fig. \ref{fig2}  where we plot the values $a_{0}$, $b_{\ast}$ and $r_{\ast}$ for the range of values of $D$. As expected, the scale factors acquire the maximum value at the de Sitter radius and shrink for increasing backreaction. The same is true for the range of the radial displacement $r^{\ast}$. Overall, the string gravity acts as to shrink the size of $S^{1}$ and $S^{2}$ parts of the metric.

\begin{figure}[t!]
		\includegraphics[scale=0.18]{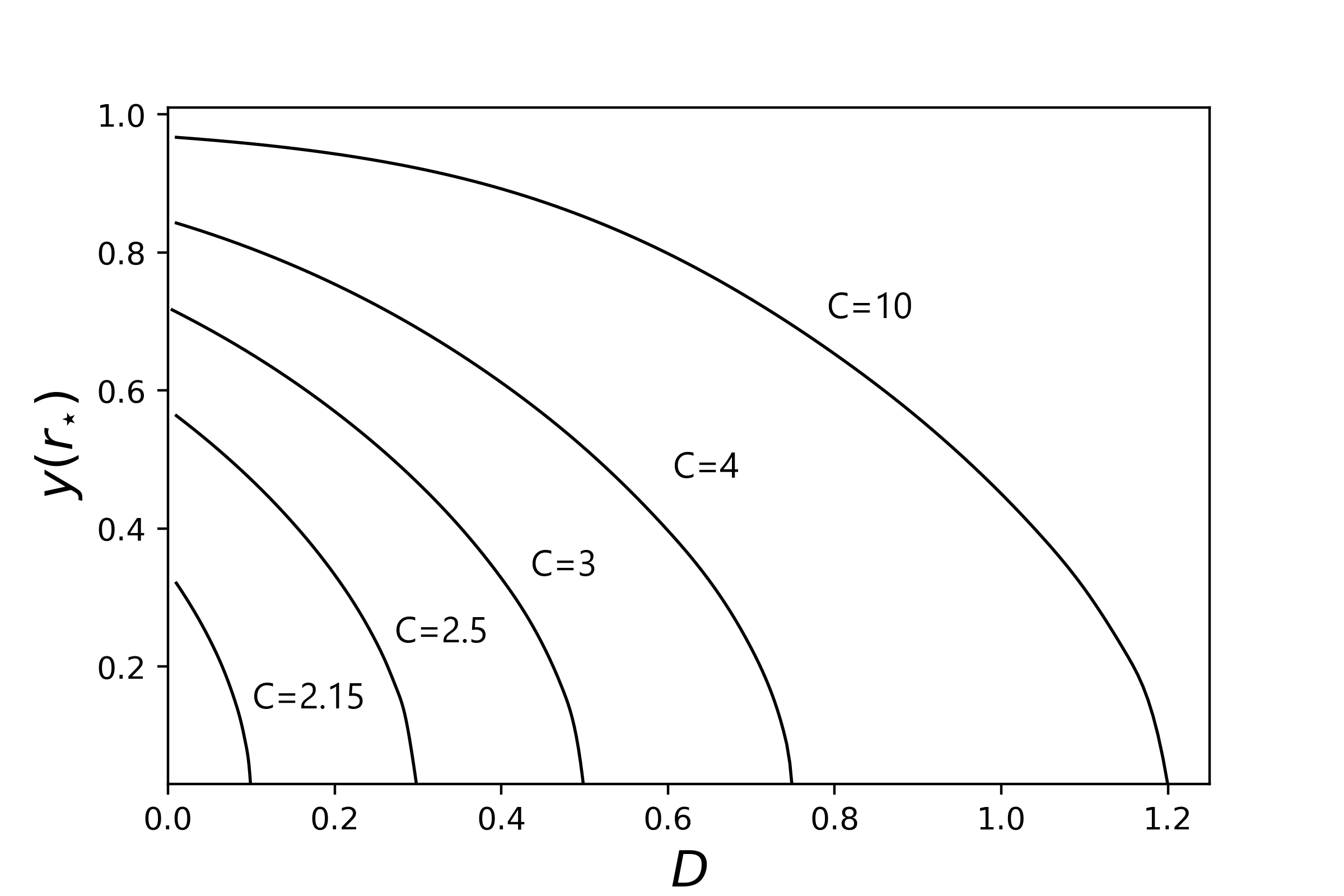}
		\caption{A plot of the field value at the horizon $y(r^{\ast})$ with respect to $D$ for different values of $C$. }.
		\label{fig3}
 
\end{figure}

Finally, we are interested in the value that the field acquires at the horizon $y(r^{\ast})$ for different parameters $C$ and $D$. As already illustrated, a thick defect results in the false vacuum covering most of Euclidean geometry inside the horizon. The same is true in the case of strong gravitational backreaction for which the location of the horizon is shifted closer to the core. In both cases we expect $y(r^{\ast})$ to be close to zero. On the other hand, in the absence of gravity and in the thin-defect limit $y(r^{\ast})$ should be at the vacuum expectation value. The results are illustrated in Fig. \ref{fig3}

\subsection{The homogeneous instanton }
\label{homogeneous-section}

As shown in the previous examples there are some thresholds for the values for the parameters $C$ and $D$ for which the field is at the maximum $y=0$ throughout the whole range of $r$. Here we will analytically study this regime and determine precisely the condition for the appearance of these homogeneous configurations. Let us start by taking equations (\ref{eq1}) and (\ref{eq2}) with the approximation $y\approx0$ we have:
\beq
2aa^{\prime\prime}+{a^{\prime}}^{2}-1+3a^{2}\left(1+\frac{CD}{3}\right)=0~,
\eeq
and
\beq
{a^{\prime}}^2+2aa^{\prime}\frac{b^{\prime}}{b}-1+3a^{2}\left(1+\frac{CD}{3}\right)=0~.
\eeq
The above can be analytically solved for the scale factors $a$ and $b$ with the solutions being:
\beq
a=\frac{\cos\tilde{r}}{\sqrt{1+\frac{CD}{3}}}~,~~b=\frac{\sin\tilde{r}}{\sqrt{1+\frac{CD}{3}}}~,
\label{distorted scale f}
\eeq
where $\tilde{r}=r\sqrt{1+\frac{CD}{3}}$ and its range is $\tilde{r}\in\left[0,\pi/2\right]$. 

We now turn our attention to scalar field equation (\ref{scalar eom}). Neglecting the quadratic term on the parenthesis of the right hand and inserting (\ref{distorted scale f}) we can arrive at:
\beq
\frac{d^{2}y}{d\tilde{r}^{2}}+\left(\frac{-2+3\cos^{2}\tilde{r}}{\cos\tilde{r}\sin\tilde{r}}\right)\frac{dy}{d\tilde{r}}-\left(\frac{1}{1-\cos^{2}\tilde{r}}-\frac{2C}{1+\frac{CD}{3}}\right)y=0~.
\eeq
Yet another change of variables $u=\cos^{2}\tilde{r}$ transforms the above the hypergeometric equation to the following form:
\beq
4u\left(1-u\right)\frac{d^{2}y}{du^{2}}+2\left(3-5u\right)\frac{dy}{du}-\left(\frac{1}{1-u}-\frac{2c}{1+\frac{cd}{3}}\right)y=0~.
\eeq
The solution that satisfies the boundary condition at the horizon (\ref{phidot at horizon}) is given up to a multiplicative factor by:
\beq
y(u)={_{2}F_{1}}\left(\alpha,\beta,3/2,u\right)\frac{1}{{\sqrt{1-u}}} ~,
\label{hyper sol}
\eeq
where $_{2}F_{1}$ is the hypergeometric function and:
\beq
\alpha,\beta=\frac{1}{4}\left(1\mp\sqrt{9+\frac{8C}{1+CD/3}}\right)~.
\eeq
Making use of the hypergeometric identities, the solution (\ref{hyper sol}) can be further expressed as:
\beq
y(u)\propto \frac{y_{1}(u)+y_{2}(u)}{\sqrt{1-u}}~,
\eeq
where:
\beq
y_{1}(u)=\frac{{_{2}F_{1}}(\alpha,\beta,3/2,1-u)}{\Gamma\left(\frac{3}{2}-\alpha\right)\Gamma\left(\frac{3}{2}-\beta\right)}
\eeq
and
\beq
y_{2}(u)=\frac{\Gamma\left(\alpha+\beta-\frac{3}{2}\right)}{\Gamma(a)\Gamma(b)}\left(1-u\right){_2F_{1}}(3/2-\alpha,3/2-\beta,2,1-u)~.
\eeq
It is clear that in order for the solution to also satisfy the boundary condition $y(r=0)=0$ at the core of the string, the first term should be absent\footnote{One might worry that the coefficient in front of the second term diverges due to the factor $\Gamma(\alpha+\beta-\frac{3}{2})$, but this is not an issue because the final solution will be properly normalized.}. This in turn means that either $3/2-\alpha=-n$ or $3/2-\beta=-n$ where $n=0,1,2,3...$. Selecting the lowest possible integer for which this is satisfied $(n=0)$ yields the constraint:
\beq
\frac{1}{C}+\frac{D}{3}=\frac{1}{2}~.
\label{c and d}
\eeq
This is precisely the condition for which homogeneous instanton appears. Neglecting gravity yields the familiar result $C=2$. We note that for $C<2$ there is no value of $D$ that satisfies (\ref{c and d}) and the same goes for $D>3/2$. This is the parameter space for which the only acceptable solution is the homogeneous deSitter instanton. 

The appearance of the homogeneous, Hawking-Moss like configuration can be given a simple physical interpetation. Let us consider the string thickness $\delta$ which in de-Sitter units is given by $\delta=1/\sqrt{C}$. We expect the false vacuum of the core to cover all of space when the core size is similar to that of the horizon:
\beq
\delta\sim b_{\ast}
\label{deltasimb}
\eeq
the expression for the horizon size $b_\ast$ when we approach the homogeneous instanton is given by setting $\tilde{r}=\pi/2$ in (\ref{distorted scale f}). Inserting in (\ref{deltasimb}) we obtain:
\beq
\frac{1}{\sqrt{C}}\sim\frac{1}{\sqrt{1+\frac{CD}{3}}}\rightarrow\frac{1}{C}+\frac{D}{3}\sim 1
\eeq
which agrees with (\ref{c and d}).

Thus, we can either increase the thickness $C$ and expand the core all the way to the horizon, or increase the backreaction $D$ and distort the geometry in a way that the horizon approaches the core.

\section{Global Structure and Interpretation}

\subsection{Global Structure of the Geometry}

As discussed previously, in certain regions of the parameter space of our model, the solutions are nearly indistinguishable from a pure de Sitter solution expressed in a specific slicing (see Appendix \ref{deSitter-appendix} for details on the pure de Sitter solution in the relevant gauge for us.).

This observation suggests that, analogous to the procedure in pure de Sitter space, one can analytically 
continue the solution to a Lorentzian signature by taking the metric in the form:
\beq
ds_{E}^2=dr^{2}+a^{2}(r)\left(-dt^{2}+\cosh^{2}t~d\chi^{2}\right)+b(r)^{2}d\theta^{2}~,
\label{metric}
\eeq
while the scalar field remains as in the Euclidean solution:
\beq
\phi = \varphi(r) e^{i\theta}
\eeq

This analytic continuation does not alter the structure of the equations of motion, thus the numerical solutions presented earlier also apply to these Lorentzian configurations.

Furthermore, as in the pure de Sitter case, this solution can be extended beyond the horizon in a smooth manner by using a metric of the form:
\beq
ds^2=H^{-2} \left[-dT^{2}+\tilde a^2(T)\left(d\psi^{2}+\sinh^{2}\psi 
~d\chi^{2}\right)+\tilde b^2(T)d\theta^{2}\right]~.
\eeq
where the function $\tilde a^2(T)$ and $\tilde b^2(T)$ can be obtained from the equations of motion but their
leading behaviour is fixed by imposing the smoothness of the horizon at $T\rightarrow 0$.
This metric represents a cosmological spacetime with anisotropic spatial slices. Two of the dimensions correspond to an infinite open universe, while the angular coordinate $\theta$ parametrizes an $S_1$ component of the geometry.
Moreover, the matter content in this universe is, once again, provided by a complex scalar field winding along the 
circle. Solutions of this type have been studied in the context of dimensional decompactification and anisotropic phases in the early universe \cite{Blanco-Pillado:2010rbj}.

It is important to note that this description of the spacetime remains valid even in the absence of a cosmological constant \cite{Gregory:1996dd,Gregory:2002tp}. As discussed earlier, the global spacetime structure of a global string in flat space is remarkably similar to the structure when a cosmological constant is included. The primary difference in these scenarios lies in the subsequent evolution of the spacetime. 

In this work, we do not numerically evolve these configurations and instead just provide a brief discussion of the 
possible types of solutions. Depending on the parameter values in the model, a range of configurations can emerge, from pure flat space, which asymptotically approaches a Bianchi-type solution, to other solutions 
exhibiting a brief period of anisotropic inflation, and extending all the way to cases where eternal inflation 
occurs in the regime of topological inflation \cite{Vilenkin:1994pv,Linde:1994wt}.

\subsection{Interpretation of the solutions}

As discussed earlier, the solutions presented in this paper allow for a dual interpretation. On the one hand, they 
can be viewed as instantons describing the nucleation of global strings in de Sitter space. In this context, 
these configurations and their subsequent evolution could be relevant to the early universe \cite{Garriga:1992nm, Conlon:2024uob}.

Moreover, even in the absence of a cosmological constant, these configurations may have cosmological 
significance for a large value of $D$. For example, one could imagine a global string produced during 
inflation, which persists into later stages of the universe's evolution, when the energy density is much lower, 
making it effectively describable by solutions with $H=0$ and obtained in Sec. 4.1 . The radial component of this 
metric resembles the solutions in the case of domain walls \cite{Deng:2016vzb} so it is likely that a wormhole-like structure will develop in these situations as well. However, the angular component complicates this interpretation. We leave further exploration of this intriguing possibility for future work.

Alternatively, these solutions can be interpreted in the framework of quantum cosmology. In this context, the 
universe could be created from "nothing," a process that can be described using gravitational instantons. The 
classic example of such creation is pure de Sitter space, where the corresponding instanton is the round Euclidean 4-sphere. Its analytic continuation results in the formation of a closed de Sitter universe \cite{Vilenkin:1983xq}.

In a more general scenario, one might consider including some form of matter content in the universe's creation 
process, such as topological defects like domain walls or strings. For instance, a model described in \cite{Fanaras:2023acz,Blanco-Pillado:2019tdf} 
considered the creation of the universe with a domain wall wrapping around its equator, referred to as a "domain wall universe". Here, we have explored the existence of instantons with global strings. 

%The computation of the Euclidean 
%action suggests that the probability of forming a universe with a string may be higher than that of forming a pure 
%de Sitter universe. Additionally, given the lower symmetry of string instantons, this seems to imply a preference 
%for the creation of anisotropic universes in quantum cosmology. Hence we expect that "global string universes" could be
%preferred in models that posses string excitations of this kind.

A particularly interesting aspect of these solutions is the interpretation of the flat-space case ($H=0$) as a 
quantum cosmology instanton. The structure of this solution indicates that one could contemplate the creation of 
a universe from nothing, even in the absence of a cosmological constant. Depending on the parameters of the model, it 
is also possible for these models to exhibit a period of anisotropic inflation in the region beyond the horizon. 
In that regard, these type of solutions give a motivation for the somewhat unnatural initial conditions of the universe 
in these models.

\section{The Euclidean Action}

\subsection{The Bounce Action}

The tunneling rate (\ref{rate}) evaluated in Sec. 1 is valid so long the the string core is sufficiently thin compared to the horizon and the gravitational effects of the coupling of the string to the Goldstone mode are negligible. A more complete treatment of the nucleation process is achieved through the evaluation of the Euclidean action (\ref{action}). The nucleation rate is given by the expression \cite{Coleman:1980aw}:
\beq
P\sim e^{-B}~,
\eeq
where:
\beq
B=S_{E}-S_E^{dS}~,
\label{Bounce}
\eeq
is the bounce action and $S_E^{dS}=-\pi/(GH^{2})$ is the Euclidean de Sitter action. The above rate can be understood as the probability to nucleate a global string characterized by the parameters $C$ and $D$ in a de Sitter background of cosmological constant $H^{2}$.

In order to evaluate the Euclidean string action, we proceed by inserting the equation of motion (\ref{eqofm}) in (\ref{action}) and perform integration. We arrive at:
\beq
S_{E}=\frac{\pi}{G}\left[\frac{d}{dr}\left(a^{2}b\right)-2aba^{\prime}\right]_{0}^{r_{\ast}}+16\pi^{2}\int_{0}^{r_{\ast}} dr\frac{a^{2}}{b}\varphi^{2}~,
\eeq
which can be recast as,
\beq
S_{E}=-\frac{\pi a_{0}^{2}}{ G}+16\pi^{2}\int_{0}^{r_{\ast}} dr\frac{a^{2}}{b}\varphi^{2}
\eeq
after imposing the instanton boundary conditions presented in subsection 3.1. Finally, in the scaled variables the 
expression for the action is:
\beq
\tilde{S}_{E}=-a_{0}^{2}+2D\int_{0}^{r_{\ast}}\frac{a^{2}y^{2}}{b}dr~,
\eeq
where $\tilde{S}_{E}=GH^{2}S_{E}/\pi$ and $a_0$ is described in terms of the Hubble scale units. We note the presence of two terms. A geometric component that captures the gravitational sector and a contribution from the field profile responsible for the string dynamics. We note that the latter is positive and, for the most part, subdominant to the gravitational contribution. The action for the homogeneous configuration can be readily found by using (\ref{distorted scale f}) to express $a_{0}$. Setting $y=0$ one arrives at:

\beq
\tilde{S}_{E}^{HM}=-\frac{1}{1+\frac{CD}{3}}
\label{HM action}
\eeq
It is evident from the above expression that the effective cosmological constant-in deSitter units- is $1+CD/3$.

Overall, we compute the bounce action: 
\beq
\tilde{B}=1-a_{0}^{2}+2D\int_{0}^{r_{\ast}}\frac{a^{2}y^{2}}{b}dr~,
\eeq
As expected, the geometric ``Gibbons-Hawking'' contributions cancel out in the limit of minimal gravitational distortions. Furthermore, since $a_{0}\leq1$ the bounce is always positive and as a result the nucleation process is exponentially suppressed. The expression of the homogeneous bounce can be readily found by using (\ref{HM action}) and (\ref{Bounce}):
\beq
\tilde{B}^{HM}=\frac{1}{1+\frac{3}{CD}}
\eeq

\begin{figure}[h]
  \begin{subfigure}[t]{0.475\textwidth}
    \includegraphics[scale=0.6]{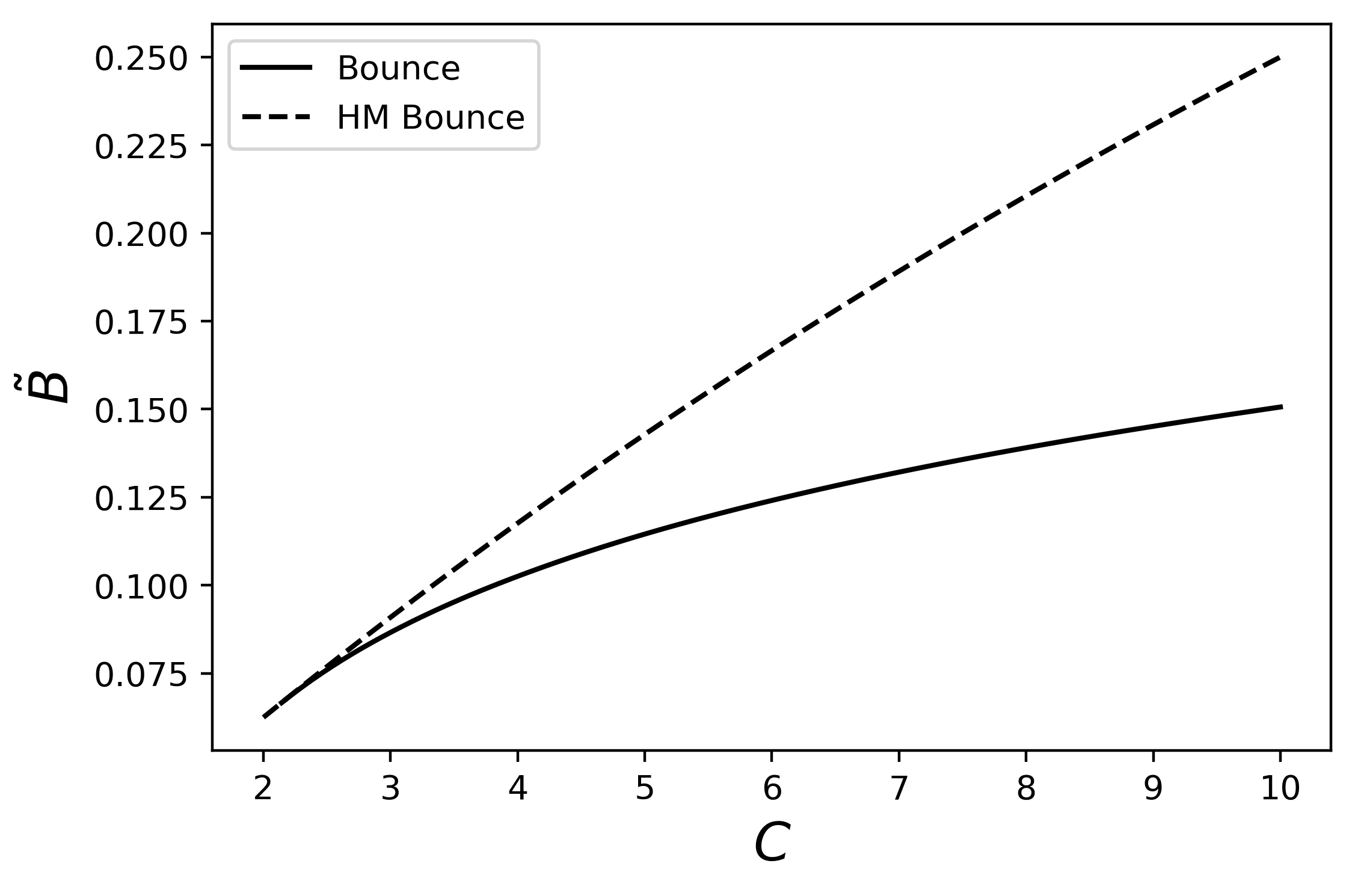}
    \label{fig-4a}
  \end{subfigure}\hfill
  \begin{subfigure}[t]{0.475\textwidth}
    \includegraphics[scale=0.6]{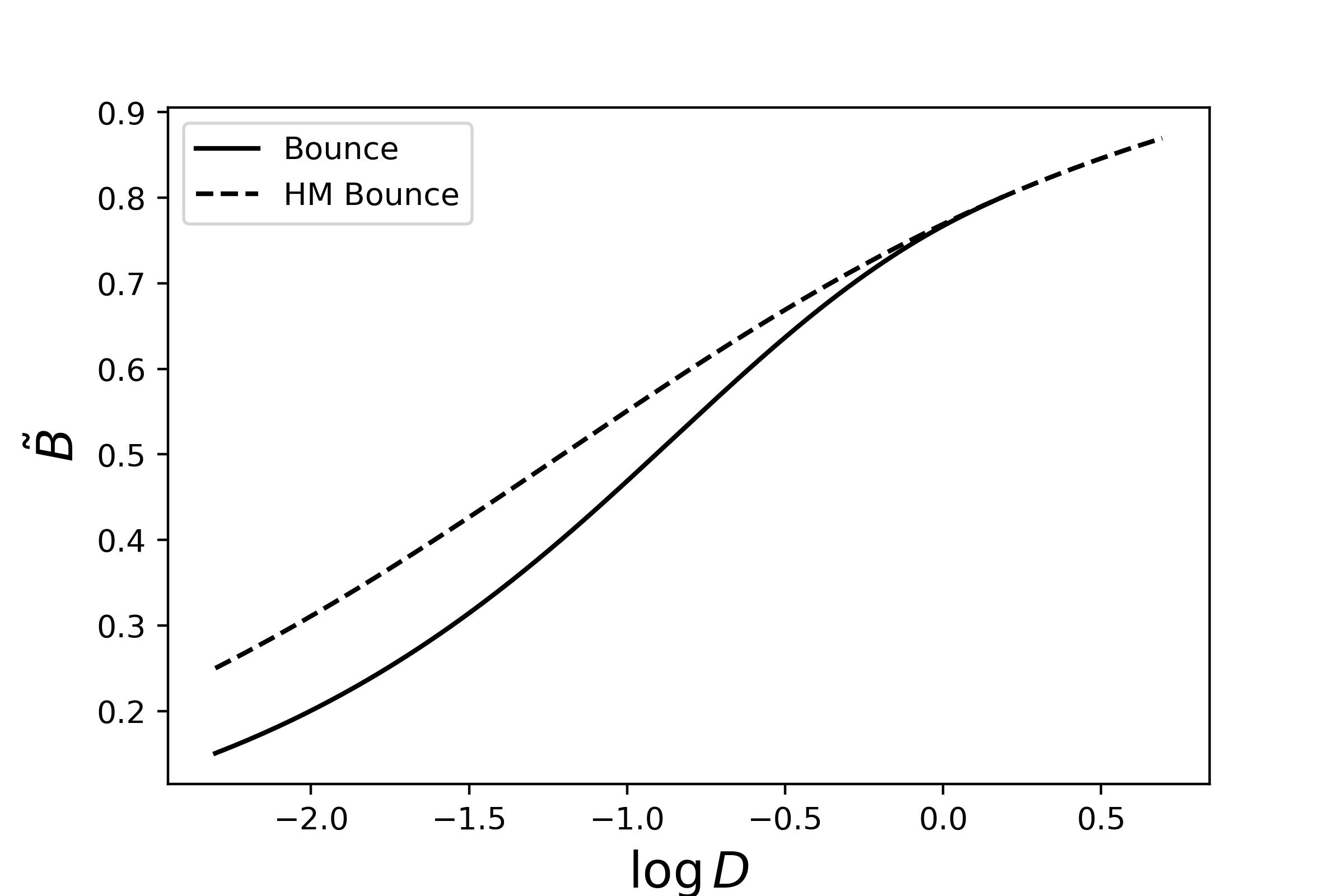}
    \label{fig-4b}
  \end{subfigure}
  \caption{Plots of the bounce action, $\tilde B$, for different values of $C$ and $D$, keeping $D=0.1$ (Left) and $C=10$ fixed (Right), respectively.} 
  \label{fig6}
\end{figure}

We plot the bounce action for several values of the parameters $C$ and $D$ in Fig.\ref{fig6}. It is clear that the nucleation rate is enhanced for thick strings with minimal backreaction. 

\subsection{The creation of the Global String Universe from Nothing}

As discussed earlier, the instantons identified in this paper can also be interpreted within the framework 
of quantum cosmology. However, while the instantons themselves remain the same, the calculation of the 
probabilities for various processes depends on the specific prescription applied.

In this work, we adopt the tunneling wavefunction prescription to compare the creation of a homogeneous 
universe with that of a "global string universe". According to this approach, the nucleation probability is given by \cite{Vilenkin:1984wp}:

\beq
P\sim e^{-|S_E|}
\eeq

\begin{figure}[h]
  \begin{subfigure}[t]{0.475\textwidth}
    \includegraphics[scale=0.6]{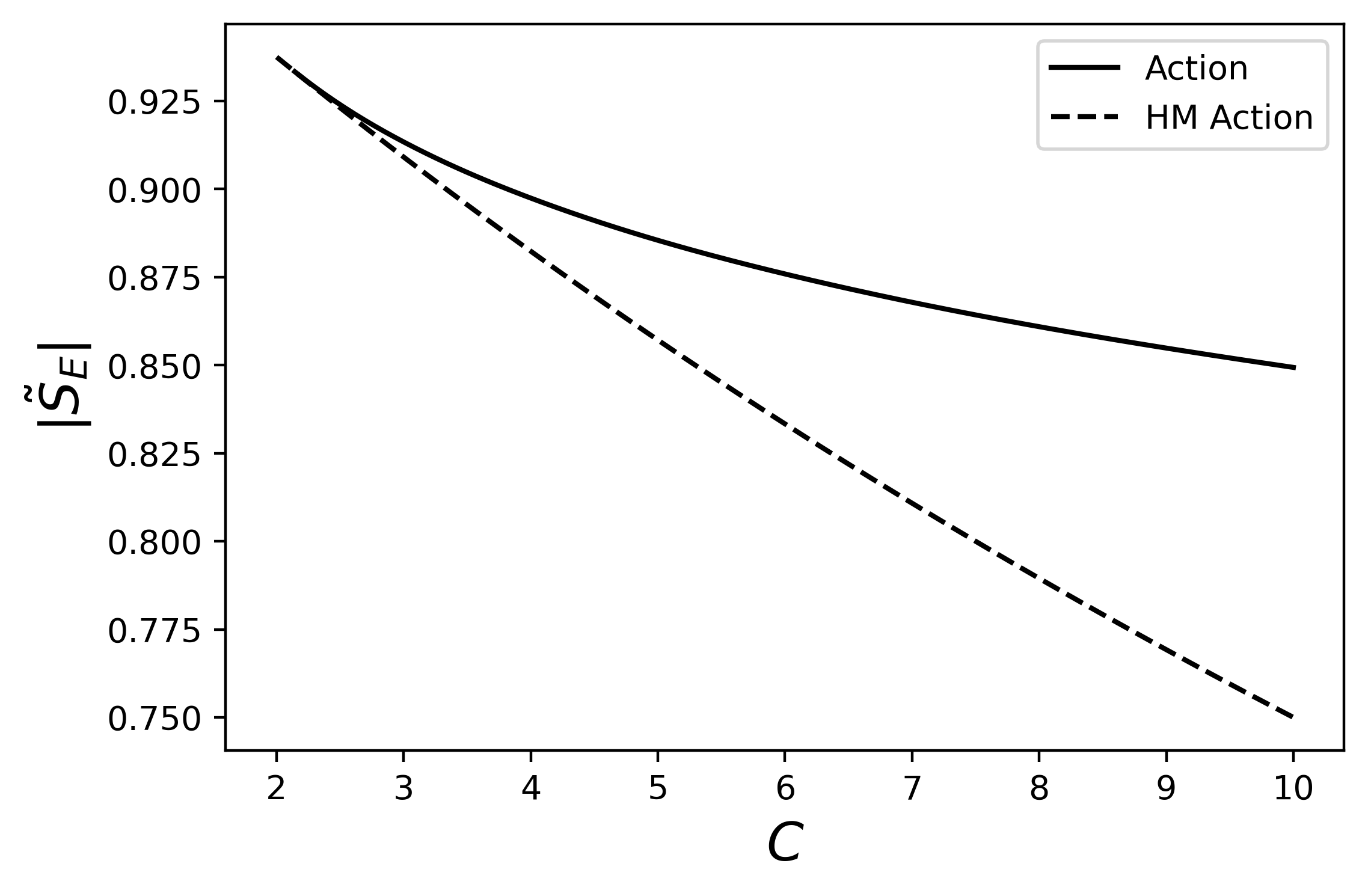}
    \label{fig-4a}
  \end{subfigure}\hfill
  \begin{subfigure}[t]{0.475\textwidth}
    \includegraphics[scale=0.6]{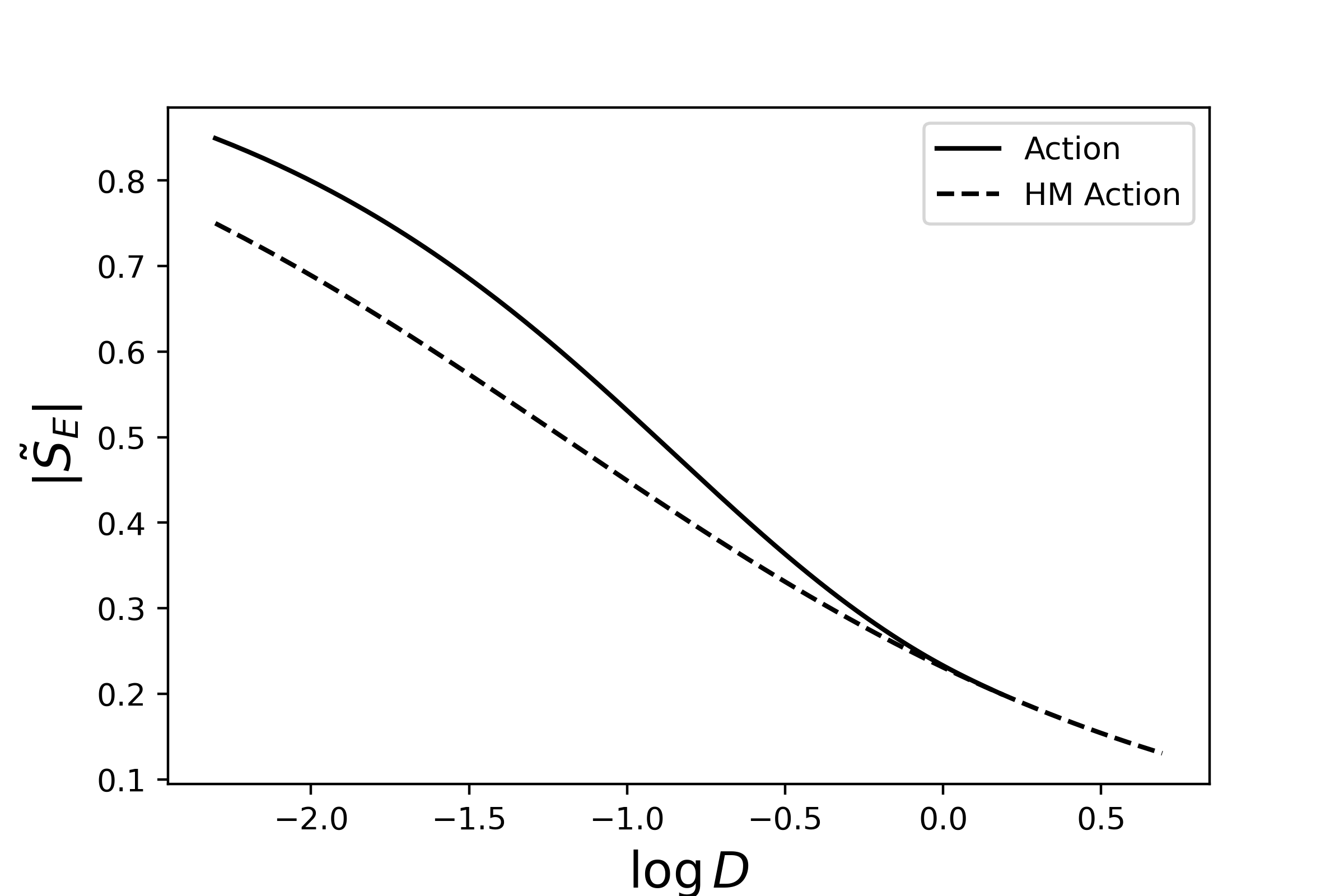}
    \label{fig-4b}
  \end{subfigure}
  \caption{Plots of the Euclidean action for different values of $C$ and $D$, keeping $D=0.1$ and $C=10$ fixed, respectively.} 
  \label{fig7}
\end{figure}

The computation of the Euclidean action suggests that the probability of forming a universe with a string may be lower 
than that of forming the homogeneous de Sitter universe at the top of the potential for the scalar field as shown in Fig.\ref{fig7}. However, 
it will be preferred with respect to the formation of a pure de Sitter universe with scale $H$. 

It is important to note, however, that this conclusion could be modified by a more detailed analysis of the quantum state of the scalar field in each of these scenarios. In particular, it was shown in \cite{Fanaras:2023acz} that the nucleation rate of a spherical universe with a scalar field governed by a hilltop potential undergoes a sharp transition when the curvature of the potential exceeds a certain threshold. This seems to suggest that the homogeneous instanton is not realized as a stable initial state for a certain regime of the parameter space $\{C,D\}$ and the inhomogeneous global string configuration is the only viable nucleation channel. We leave the exploration of this subtlety for future work.

\section{Conclusions}

In this work we provided a complete treatment of the global string instantons. Strings along with other topological defects can spontaneously nucleate at a continuous rate during the inflationary period and as such reach appreciable number densities in our local horizon. The nucleation process occurs as quantum mechanical tunneling of a defect into a horizon size object that is later stretched by the expansion of the universe. The rate for such a process is governed by the defect instanton which is a two parameter function: the core thickness relative to the horizon, $C^{-1}$,  and the gravitational backreaction on the de Sitter background geometry, $D$. The instanton analysis has been carried out in \cite{Basu:1991ig,Basu:1992ue} in the case of domain walls, while strings and monopoles were only studied in the analytically accessible limits. Here we continued this work and provide a complete treatment of the global string instantons for the whole range of relevant parameters $C$ and $D$.

We initiate our analysis by considering a complex scalar field, $\phi$, with a "Mexican hat" potential, representing a global string loop embedded in an Euclidean background spacetime. The instanton equations of motion, together with the Hamiltonian constraint, are derived for the field profile $\varphi$ and the scale factors $a$ and $b$ that parametrize the metric. We then numerically solve this system across a wide range of parameters of the underlying model, ensuring regularity conditions that guarantee a smooth 4-geometry. The boundary conditions governing these solutions turns these equations into a boundary value problem, which can be addressed using multiple shooting methods.

%The system, however, does not allow a direct application of shooting from one boundary to the other, since both boundaries exhibit numerical singularities. We were able to remedy this complication by expanding the solution in each end-point and integrating towards a specified midpoint of the integration range where the solution is regular. This method belongs in the class of multiple shooting, which addresses the issue of computational error acquired during integration, by splitting the integration interval into sub-intervals.  

We present results for the field profile $\varphi(r)$ and the scale factors $a(r)$ and $b(r)$ across different string core thicknesses. While the core size does not significantly alter the overall geometry, it does influence the range of values for the field profile $\varphi$. We also examine the impact of varying gravitational backreaction. When gravitational effects become significant, the resulting distortion of the geometry causes the horizon to shift closer to the string core. In both cases, we identify a regime in which the potential's maximum extends across the entire Euclidean space up to the horizon, resembling a homogeneous Hawking-Moss configuration.

We also discuss the analytic continuation of these solutions to Lorentzian signature. Our analysis reveals that the solutions describe global string loops expanding with an induced metric corresponding to a 2-dimensional de Sitter space. This result is in agreement with the nucleation process described in the thin wall approximation. Additionally, the global structure of these gravitational instantons can be extended beyond the horizon of the Euclidean geometry. Unlike the open universe geometries typical of Coleman-de Luccia (CdL) instantons, the creation of these global strings leads to the formation of a region with the structure of an anisotropic FRW universe. This anisotropy arises due to the reduced symmetry of the instanton caused by the presence of the string, resulting in distinct initial conditions. Beyond the horizon, the universe exhibits an open $2+1$ dimensional structure with an expanding compact dimension wound by the phase of the scalar field present in our model.

Interestingly, these solutions bear similarities to other configurations found in the literature, particularly in the context of extra dimensions \cite{Olasagasti:2000gx,Gregory:1999gv,Cho:2003gn}. Specifically, these instanton solutions can be seen as 4-dimensional analogs of the "Bubbles from Nothing" scenarios discussed in the framework of flux compactifications \cite{Blanco-Pillado:2011fcm}. In those models, the solution terminates on an expanding brane, which in our case corresponds to the vortex core. Following this connection with higher dimensional solutions, it would be interesting to explore whether similar solutions exist in spacetimes with a negative cosmological constant that resemble the ones obtained already in the litearature (See for example \cite{Gregory:1999gv} or \cite{Blanco-Pillado:2010xww}).

%\textbf{(This needs work)}We finally comment on the instanton solutions obtained and their relevance in the context of quantum cosmology. The nucleation process describes the creation of loops of string in a background deSitter space. However, the same instanton describes the creation of a universe out of nothing with a closed global string in it. Here "nothing" is a state of no classical spacetime. It was shown in [] that given a multifield potential with a maximum the homogeneous configuration will be the dominant nucleation channel as long as $V''/V<2$. For larger values of steepness and depending on the number of field components of the underlying theory the nucleation process yields a domain wall universe ,for 1-field, a global string or a global monopole. Thus, it seems that the fact the the string action is less then the homogeneous is irrelevant since the latter becomes unstable when the former channel appears. The two instantons are a natural continuation of eachother. We note that we leave the case of nucleation of monopole-antimonopole pairs for future work, but we expect the qualitative analysis to follow that of domain wall and global string nucleation. 

Finally, we also comment on the possible implications of the instantons described in this work to
Quantum Cosmology. Similar to the nucleation of strings in de Sitter space, these instantons also
describe the creation of an anisotropic universe from nothing. This suggests a connection between
the presence of the axionic Goldstone mode and the creation of anisotropic universes. In this context,
the solutions provide a natural explanation for the specific initial conditions in these spacetimes, offering some
insight into the origin of such anisotropies which otherwise may seem arbitrary.

\begin{acknowledgements}
We are grateful to Jaume Garriga, Asier Lopez-Eiguren, Ken D. Olum and Oriol Pujolas for stimulating discussion.  G. F. acknowledges support from the Constantine and Patricia Mavroyannis fellowship and the John F. Burlingame fellowship. J.J.B.-P. has been supported in part by the PID2021-123703NB-C21 
grant funded by MCIN/AEI/10.13039/501100011033/and by ERDF;“ A way of making Europe”; 
the Basque Government grant (IT-1628-22) and the Basque Foundation for Science (IKERBASQUE).
\end{acknowledgements}

\appendix

\section{Pure de Sitter space instanton}
\label{deSitter-appendix}

\subsection{Euclidean de Sitter metric}

In this Appendix we would like to describe the $4d$ Euclidean de Sitter space in a gauge which is particularly useful for the comparison with our instantons. As is well known 4d-Euclidean de Sitter space can be described as a $4d$ sphere of radius $H^{-1}$ by embedding it in a 5-dimensional Euclidean space. Here we would like to do this by using the following coordinate system
\beqa
t &=& H^{-1} \cos(r) \sin(\tau)~, \\
w &=& H^{-1} \cos(r) \cos(\tau) \cos(\chi)~, \\
z &=& H^{-1} \cos(r) \cos(\tau) \sin(\chi)~, \\
x &=& H^{-1} \sin(r) \cos(\theta)~, \\
y &=& H^{-1} \sin(r) \cos(\theta)~,
\eeqa
which clearly satisfies the condition $t^2+w^2+x^2+y^2+z^2=H^{-2}$. Therefore, this coordinate system represents a particular chart of a 
$4d$ Euclidean sphere with an induced metric of the form,
\beq
ds_{E}^2=H^{-2} \left[dr^{2}+\cos^2(r)\left(d\tau^{2}+\sin^{2}\tau 
~d\chi^{2}\right)+\sin^2(r)d\theta^{2}\right]~.
\label{metric-puredS}
\eeq

We note that the form of this metric is consistent with the general ansatz we use to parametrize our instanton solution in Eq. (\ref{metric}). 
Moreover, as this metric corresponds to pure de Sitter space, it is manifestly regular, and the boundary conditions are fully consistent with those imposed to ensure a smooth geometry in our general instanton solutions, as specified in Eqs. (\ref{bc1},~\ref{bc2},~\ref{bc3} ).

We adopt this slicing of de Sitter space as the reference metric for all global string instantons considered in this work. As demonstrated in the main text, there exist several distinct regions in the parameter space where the solutions closely approximate the pure de Sitter metric presented here. In other cases, while the geometry exhibits greater distortion, it retains its overall structure and satisfies the same boundary conditions.

\subsection{Lorentzian continuation}

We can analytically continue this metric to the Lorentzian signature to obtain a
slicing of de Sitter space of the form,

\beq
ds^2=H^{-2} \left[dr^{2}+\cos^2(r)\left(-dt^{2}+\cosh^{2}t 
~d\chi^{2}\right)+\sin^2(r)d\theta^{2}\right]~.
\eeq

This particular chart of de Sitter space exhibits a notable structure. If we momentarily disregard the angular part of the metric, the $d\theta$ part, it becomes apparent that the geometry closely resembles the Coleman-de Luccia (CdL) ansatz for a domain wall universe, but in a lower dimensional case, in 2+1 dimensions. Based on this, one might expect our $4d$ solution to exhibit two distinct horizons at the extremes of the radial coordinate as in the CdL geometry. However, in our 4-dimensional spacetime, this is not the case.

Specifically, while the geometry can indeed be extended beyond the region $r=\pi/2$ region, it terminates at the opposite end where the part of the geometry smoothly pinches off. This implies that our metric can only be extended beyond the horizon at $r=\pi/2$. The form of de Sitter space in this
case can be written as
\beq
ds^2=H^{-2} \left[-dT^{2}+\sinh^2(T)\left(d\psi^{2}+\sinh^{2}\psi 
~d\chi^{2}\right)+\cosh^2(T)d\theta^{2}\right]~.
\eeq
which can be identify as cosmological but anisotropic slicing of de Sitter space \footnote{One can recognize the Bianchi III form of this metric representing an initially anisotropic de Sitter space geometry. This type of geometry has been discussed in relation to anisotropic quantum tunneling in \cite{Blanco-Pillado:2010rbj}.}. As we discuss in the main
part of the text, the instanton solutions we found in our field theory models with a global string have a similar
global structure to the one explain here for the pure de Sitter space.

\section{Numerical integration methods}
\label{numerics}

The set of equations (\ref{3x3system1})-(\ref{3x3system3}) will have to be integrated numerically for different values of the parameters $C$ and $D$. Due to the nature of the boundary conditions the task belongs in the category of boundary value problems and a solution can be found via shooting methods. The general idea is to reduce the boundary value problem to an initial value one by guessing the values of the variables at one boundary, integrating and checking whether the boundary conditions at the other boundary are met. This procedure can be automated by utilizing optimization methods to determine the value of the shooting parameter for which the conditions at each boundary are satisfied.

A non-trivial characteristic of our system is the presence of two shooting parameters, $a_{0}$ and $y^{\prime}_{0}$, if we integrate from the core of the string, and $b_{\ast}$ and $y_{\ast}$ if we start from the horizon. Thus, we will have to resort to a multivariable optimization method such as Powell's hybrid method in order to determine the appropriate boundary values. We also note that since our system is autonomous we are free to integrate from either side while making the appropriate rescaling of the $r$-coordinate.

A further complication is the fact that both boundaries exhibit potential numerical singularities. Such singularities are not present analytically, but arise numerically at the boundaries of integration due to the presence of terms sensitive to large numerical cancelations. As a result shooting from one side is a highly unstable process. To better understand this obstacle consider that we pick the values of the shooting parameters at the horizon with a reasonably good precision and shoot towards the core. While the code runs it will gradually pick up computational error. Eventually, we want the integration to terminate once the boundary conditions are met. In particular this means once the condition $b_{0}\rightarrow0$ and $y\rightarrow0$ is met with an acceptable accuracy. For equation (\ref{scalar eom}) to yield consistent results we should expect a cancellation of the terms $b^{\prime}y^{\prime}/b-y/b^{2}$ which go symbolically as $\sim 1/0-1/0$. This is highly sensitive to the error accumulated during the calculation because even if $b$ and $y$ approach minuscule values, they might nevertheless differ by a few orders of magnitude. This results in the appearance of, at least, $\mathcal{O}(1)$ order terms which analytically should not be present. The same phenomenon is also present if we shoot from the core to the horizon. Thus, we need to figure a method that does not run into this issue.

Taking all this into consideration, we implemented a double shooting method which belongs in the broader category of multiple shooting \footnote{For a thorough analysis on multiple shooting methods in calculating tunneling rates see \cite{Masoumi:2016wot}.}. The motivation behind multiple shooting is to split the integration interval in many sub-intervals so as to control the accumulation of error which depends exponentially on the amount of integration. The only compromise is that the shooting parameters multiply by the number of intervals we choose to split the initial range. In our case, we find it sufficient to shoot from both sides and match the solutions at an intermediate point. We first Taylor expand the equations near each boundary in order to deal with the singular behavior that arises in numerical integration. Using a Runge-Kutta method of order 8, denoted in Python as $DOP853$ we shoot towards one end and terminate once the scale factor $a$ reaches $\approx0.4$, a value which is roughly half $a_{0}$ and sufficiently far away from each boundary. We then shoot from the latter boundary and terminate at the exact same point, which is not plagued by numerical singularities. The precision of our calculation is measured by taking the differences of all variables and their first and second derivatives at the meeting point.  The shooting parameters at the core $a_{0}$, $y^{\prime}_{0}$ and at the horizon $b_{\ast}$, $y_{\ast}$ are then determined by solving a $4\times 4$ system at the midpoint using Powell's hybrid optimization method. Once the system has been solved, we are also able to obtain the location of the horizon $r^{\ast}$.

\end{document}